\documentclass[12pt]{article}
\usepackage{epsfig,amsfonts,amssymb}

\usepackage{comment}
\usepackage{cite}
\usepackage{cite}

\usepackage[utf8]{inputenc}
\usepackage[T1]{fontenc}
\usepackage{lmodern, textcomp}
\usepackage[english]{babel}
\usepackage{csquotes}

\usepackage[left=1in,right=1in,top=.8in,bottom=.8in]{geometry}

\usepackage{eurosym, xspace}
\usepackage{graphicx, subfig, float}
\usepackage{amsmath, amsfonts, amssymb}
\usepackage{mathtools}
\usepackage{slashed}

\usepackage[usenames, dvipsnames]{xcolor}
\usepackage[amsmath, hyperref, framed]{}%{ntheorem}
\usepackage{hhline}
\usepackage{bm}
\usepackage{array, caption}
\usepackage{array,multirow}

\newcolumntype{C}[1]{>{\centering\arraybackslash}m{#1}}

\usepackage[pdftex, breaklinks=true, linktocpage,
	colorlinks=true, urlcolor=blue, linkcolor=blue, citecolor=red
]{hyperref}

\usepackage{cite}

\def\ZZZ{{\hbox{ Z\kern-1.6mm Z}}}
\def\RRR{{\hbox{ R\kern-2.4mm R}}}
\def\CCC{{\hbox{ C\kern-2.0mm C}}}
\def\zzz{{\hbox{z\kern-1mm z}}}

\newcommand{\qeq}{{\hbox{=\kern-2.3mm ? \kern.5mm }}}
\renewcommand{\qeq}{=}

\newcommand{\vp}{\varphi}

\newcommand{\BB}{{\cal B}}

\newcommand{\AAA}{{\cal A}}
\newcommand{\GG}{{\cal G}}

\newcommand{\HH}{{\cal H}}
\newcommand{\MM}{{\cal M}}

\newcommand{\OO}{{\cal O}}

\newcommand{\PP}{{\cal P}}

\newcommand{\XX}{{\cal X}}

\newcommand{\wt}{\widetilde}
\newcommand{\wh}{\widehat}

\newcommand{\RR}{{\cal R}}

\newcommand{\be}{\begin{equation}}
\newcommand{\ee}{\end{equation}}
\newcommand{\ben}{\begin{eqnarray}\displaystyle}
\newcommand{\een}{\end{eqnarray}}

\newcommand{\refb}[1]{(\ref{#1})}
\newcommand{\p}{\partial}
\newcommand{\sectiono}[1]{\section{#1}\setcounter{equation}{0}}

\def\one{{\hbox{ 1\kern-.8mm l}}}
\def\zero{{\hbox{ 0\kern-1.5mm 0}}}

\newcommand{\bea}[1]{\begin{eqnarray}\label{#1} }
\newcommand{\eea}{\end{eqnarray}}

\usepackage{bm}
\newcommand\non{\nonumber}

\newcommand\f{\frac}

\def\figone{

\def\JPicScale{0.8}
\ifx\JPicScale\undefined\def\JPicScale{1}\fi
\unitlength \JPicScale mm
\begin{picture}(135,80)(0,0)
\linethickness{0.3mm}
\multiput(40,80)(0.12,-0.18){167}{\line(0,-1){0.18}}
\linethickness{0.3mm}
\multiput(30,70)(0.18,-0.12){167}{\line(1,0){0.18}}
\linethickness{0.3mm}
\put(30,50){\line(1,0){30}}
\linethickness{0.3mm}
\multiput(30,30)(0.18,0.12){167}{\line(1,0){0.18}}
\linethickness{0.3mm}
\multiput(40,20)(0.12,0.18){167}{\line(0,1){0.18}}
\linethickness{0.3mm}
\put(60,50){\line(1,0){40}}
\linethickness{0.3mm}
\multiput(100,50)(0.12,0.18){167}{\line(0,1){0.18}}
\linethickness{0.3mm}
\multiput(100,50)(0.18,0.12){167}{\line(1,0){0.18}}
\linethickness{0.3mm}
\put(100,50){\line(1,0){30}}
\linethickness{0.3mm}
\multiput(100,50)(0.18,-0.12){167}{\line(1,0){0.18}}
\linethickness{0.3mm}
\multiput(100,50)(0.12,-0.18){167}{\line(0,-1){0.18}}

\put(30,80){\makebox(0,0)[cc]{$\zeta Q_B A_1^c$}}

\put(25,70){\makebox(0,0)[cc]{$A_2^c$}}

\put(25,50){\makebox(0,0)[cc]{$A_n^c$}}

\put(20,30){\makebox(0,0)[cc]{$A_1^o$}}

\put(35,20){\makebox(0,0)[cc]{$A_p^o$}}

\put(35,60){\makebox(0,0)[cc]{$\vdots$}}

\put(40,30){\makebox(0,0)[cc]{$\vdots$}}

\put(125,80){\makebox(0,0)[cc]{$B_1^c$}}

\put(135,70){\makebox(0,0)[cc]{$B_2^c$}}

\put(135,50){\makebox(0,0)[cc]{$B_m^c$}}

\put(135,30){\makebox(0,0)[cc]{$B_1^o$}}

\put(120,15){\makebox(0,0)[cc]{$B_q^o$}}

\put(115,35){\makebox(0,0)[cc]{$\vdots$}}

\put(120,57){\makebox(0,0)[cc]{$\vdots$}}

\put(66,47){\makebox(0,0)[cc]{$\psi_r^c\vp_r$}}

\put(93,47){\makebox(0,0)[cc]{$\psi_s^c\vp_s$}}

\end{picture}

}

\def\figtwo{

\def\JPicScale{0.8}
\ifx\JPicScale\undefined\def\JPicScale{1}\fi
\unitlength \JPicScale mm
\begin{picture}(135,80)(0,0)
\linethickness{0.3mm}
\multiput(40,80)(0.12,-0.18){167}{\line(0,-1){0.18}}
\linethickness{0.3mm}
\multiput(30,70)(0.18,-0.12){167}{\line(1,0){0.18}}
\linethickness{0.3mm}
\put(30,50){\line(1,0){30}}
\linethickness{0.3mm}
\multiput(30,30)(0.18,0.12){167}{\line(1,0){0.18}}
\linethickness{0.3mm}
\multiput(40,20)(0.12,0.18){167}{\line(0,1){0.18}}
\linethickness{0.3mm}
\put(60,50){\line(1,0){40}}

\put(30,80){\makebox(0,0)[cc]{$\zeta Q_B A_1^c$}}

\put(25,70){\makebox(0,0)[cc]{$A_2^c$}}

\put(25,50){\makebox(0,0)[cc]{$B_m^c$}}

\put(20,30){\makebox(0,0)[cc]{$A_1^o$}}

\put(35,20){\makebox(0,0)[cc]{$B_q^o$}}

\put(35,60){\makebox(0,0)[cc]{$\vdots$}}

\put(40,30){\makebox(0,0)[cc]{$\vdots$}}

\put(66,47){\makebox(0,0)[cc]{$\psi_r^c\vp_r$}}

\put(93,45){\makebox(0,0)[cc]{$\tilde\psi_s^c\tilde\vp^s$}}

\put(100,50){\makebox(0,0)[cc]{$\times$}}

\end{picture}

}

\begin{document}

\baselineskip 24pt

\begin{center}
{\Large \bf   Superstring  Field Theory with Open and Closed Strings}

\end{center}

\vskip .6cm
\medskip

\vspace*{4.0ex}

\baselineskip=18pt

\begin{center}

{\large 
\rm Seyed Faroogh Moosavian$^a$, 
Ashoke Sen$^{b}$ and Mritunjay Verma$^{c}$ }

\end{center}

\vspace*{4.0ex}

\centerline{\it \small $^a$ Perimeter Institute for Theoretical Physics,  
Waterloo, 
ON N2L 2Y5, Canada}
\centerline{\it \small $^b$Harish-Chandra Research Institute, HBNI,
Chhatnag Road, Jhusi,
Allahabad 211019, India}
\centerline{ \it \small $^c$INFN, Sezione di Napoli, Complesso 
Universitario di Monte S. Angelo ed. 6}
\centerline{ \it \small via Cintia, 80126, Napoli, Italy.}

\vspace*{1.0ex}
\centerline{\small E-mail:  sfmoosavian@perimeterinstitute.ca, sen@hri.res.in,
mverma@na.infn.it }

\vspace*{5.0ex}

\centerline{\bf Abstract} \bigskip

We construct Lorentz invariant and gauge invariant 
1PI effective action for closed and open superstrings and demonstrate that it satisfies the
classical BV master equation. We also construct the quantum master action for
this theory satisfying the quantum BV master equation and generalize the
construction to unoriented theories. The extra free field needed for the construction
of closed superstring field theory plays a crucial role in coupling the closed strings to
D-branes and orientifold planes.

\vfill

%\begin{center}
%{\Huge Preliminary version}
%\end{center}

\vfill \eject

\baselineskip18pt

\tableofcontents

\sectiono{Introduction and summary} \label{s1}

We now have a consistent formulation of closed superstring field theory 
(see \cite{1703.06410}
for a review). Our goal in this paper is to extend this construction to interacting theory
of open and closed strings. Such systems arise naturally in the presence of D-branes.

As has been described in \cite{1703.06410}, full quantum 
string field theory can be studied at different
levels. The basic formulation involves the quantum master action satisfying quantum BV
master equation\cite{bv1,bv,henn}. However, without giving up any information, 
one can have various equivalent formulations that are useful
for specific studies. One of them is the 1PI effective action, suitable for studying the 
problem of mass renormalization, vacuum shift and computing S-matrix in the
shifted vacuum. Another is the Wilsonian effective action, suitable for studying the 
dynamics of string states below certain mass scale, but without making any
low energy approximation. 
On the other hand the quantum master action is useful for studying problems that
require making all loop momenta integration manifest, {\it e.g.} 
study of unitarity\cite{1604.01783} and analyticity\cite{1810.07197} properties of 
the amplitudes.
Of these the 1PI effective action has the simplest gauge transformation properties,
in that it should be invariant under suitable gauge transformation. 
As a consequence of this, it
satisfies the classical BV master equation\cite{bv1}. In contrast the quantum BV  master action and
the Wilsonian effective action are not invariant under any gauge transformation, since the
gauge non-invariance of the action needs to compensate for the gauge
non-invariance of the path integral measure\cite{bv1,bv,henn}. What remains invariant is
the combination $d\mu\, e^{2S}$ where $d\mu$ is the integration measure in the space of
fields and anti-fields and $S$ is the action\cite{9309027}.
This is ensured by the fact that the action
satisfies quantum BV master equation.

For this reason we begin our study by first constructing the 1PI effective action. 
To experts in quantum field theory, it may appear strange that we construct the 1PI effective
action before writing down the actual action whose off-shell 1PI amplitudes would
give the interaction terms of the 1PI effective action. However as explained in 
\cite{1703.06410}, since in string theory we already know the formal expressions for on-shell 
amplitudes (ignoring effects of vacuum shift and mass renormalization), we can first generalize
this to construct off-shell amplitudes, given as integrals over moduli spaces of punctured Riemann
surfaces. By appropriately restricting the region of integration over the moduli spaces, we can construct
either 1PI amplitudes or amplitudes corresponding to elementary vertices of the quantum master
action. The former requires us to remove certain regions of integration around
separating type degenerations in the moduli space, whereas the latter requires us to remove 
certain regions of integration around all degenerations.

After constructing the 1PI effective action and checking its desired properties, 
namely gauge invariance and validity of classical BV master equation, we also 
describe the construction of quantum master action satisfying the quantum BV master
equation. This construction only requires a few changes from its 1PI counterpart.
We also generalize our results to unoriented string theory, required to describe
the interacting theory of closed and open strings in the presence of orientifold planes.
Throughout our work, 
we shall follow the general strategy that has been used in the formulation of bosonic
string field theory\cite{9206084,zwiebach_open,9705241}, but extending this to superstring field 
theory requires a few additional ingredients that we shall explain. 

We now give a summary of our result for the 1PI effective action of superstring field theory
of open and closed strings. We
define $\HH_{m,n}$ to be the vector space of GSO even 
closed string states $|s\rangle$, carrying (left, right) picture numbers
$(m,n)$ and satisfying the  constraints\cite{9206084}:\footnote{These restrictions on the off-shell 
closed string states are needed to get well-defined off-shell
amplitudes, which require us to choose local coordinates at the punctures that are used to insert
the vertex operators of the off-shell string states into the correlation function. Such choice of local
coordinates at the punctures are possible globally only if we  ignore the phase of the local
coordinate. The requirement that the vertex operators must be invariant under such phase rotations
leads to \refb{eclosedcon}.
}
\be\label{eclosedcon}
b_0^-|s\rangle=0, \quad L_0^-|s\rangle=0, \quad L_0^\pm \equiv L_0\pm \bar L_0, 
\quad b_0^\pm \equiv b_0\pm \bar b_0, \quad c_0^\pm
\equiv {1\over 2}( c_0\pm \bar c_0)\, .
\ee
We also denote by $\HH_m$ the vector space of GSO even
open string states of picture number $m$.
We now introduce two subspaces in the
closed string  Hilbert space and  two subspaces in open string Hilbert space as follows:
\ben \label{edeffcho}
&& \HH^c \equiv \HH_{-1,-1}\oplus \HH_{-1/2,-1}\oplus \HH_{-1,-1/2}\oplus \HH_{-1/2, -1/2},\nonumber \\
&& \wt\HH^c \equiv \HH_{-1,-1}\oplus \HH_{-3/2,-1}\oplus \HH_{-1,-3/2}\oplus \HH_{-3/2, -3/2},\nonumber \\
&& \HH^o \equiv \HH_{-1} \oplus \HH_{-1/2}, \nonumber \\
&& \wt\HH^o \equiv \HH_{-1}\oplus \HH_{-3/2}\, .
\een
For $A^c_i\in\HH^c$ and $A^o_i\in\HH^o$, 
we define
$\{ A^c_1\cdots A^c_N; A^o_{1} \cdots A^o_{M}\}$
to be the
off-shell 1PI amplitude, summed over all genera and all number of boundaries, with 
external closed string states $A^c_1,\cdots, A^c_N$ and external open string states 
$A^o_{1}, \cdots A^o_{M}$, {\em excluding the one point function of closed string
states on the disc from the definition of $\{A^c;\}$}. In computing this amplitude, we also need to insert in the
correlation function appropriate combination of picture changing operators (PCO) and ghosts.
For
$\wt A^c\in\wt\HH^c$, we also define $\{\wt A^c\}_D$ to be the disc one point function of 
$\tilde A^c$ with
appropriate insertion of PCOs and $c$-ghosts. More detailed definitions of
$\{ A^c_1\cdots A^c_N; A^o_{1} \cdots A^o_{M}\}$ and $\{\wt A^c\}_D$ 
have been given in \S\ref{s2}.

Note the exclusion of the one point function on the disc from the definition of $\{A^c;\}$. If $A^c$ is in the
RR sector, then the contribution to $\{A^c;\}$ from the disc does not exist, 
since the RR vertex operators carry total picture number
$-1$, and therefore it is impossible to satisfy picture number conservation on the disc, requiring
total picture number $-2$, by inserting PCOs which carry positive picture number. For $A^c$ belonging to the
NSNS sector the disc one point function is non-vanishing, but we shall still exclude its contribution from the
definition of $\{A^c;\}$ for uniformity and define $\{\wt A^c\}_D$ to contain contributions from one point
function on the disc both for NSNS and RR sector $\wt A^c$. As we shall comment at the end of this section, 
whether to include
the contribution to the disc one point function of NSNS sector 
closed string states into the definition of $\{A^c;\}$ or
$\{\wt A^c\}_D$ is a matter of convention, but for RR states there is no such option. For one point function
of closed string states in $\HH^c$ on surfaces with more boundaries / genera, there is no problem with picture number
conservation, and we include their contribution in the definition of $\{A^c;\}$.

We shall take the closed string states $A^c_i$'s and $\wt A^c$ 
to be even elements of the grassmann algebra
and the open string states $A^o_i$'s and  $\wt A^o_i$'s to 
be odd elements of the grassmann algebra. 
This means that if $\zeta$ is an odd $c$-number
element of the grassmann algebra, then we have
\be\label{evenodd}
\zeta A^c_i=A^c_i\zeta, \quad \zeta \wt A^c = \wt A^c \zeta, \quad \zeta A^o_i=-A^o_i\zeta, \quad 
\zeta \wt A^o_i=-\wt A^o_i\zeta\, ,
\ee
where in \refb{evenodd}, $A^c_i$, $\wt A^c$, $A^o_i$  and $\wt A^o$ 
refer to the vertex operators corresponding to
the states.\footnote{Ref.\cite{9705241} assigned even grassmann parity to the open string field by taking the
open string vacuum $|0\rangle$ to have odd grassmann parity and identified the grassmann parity of the
string field as that of the corresponding ket state. In contrast we take the grassmann parity of the string
field to that of the corresponding vertex operator.}
Unless mentioned otherwise, all our subsequent equations will assume this assignment of
grassmann parities. In this case,
$\{ A^c_1\cdots A^c_N; A^o_{1} \cdots A^o_{M}\}$ and $\{\wt A^c\}_D$ can be shown to be
even elements of the grassmann algebra.
Furthermore it follows from the definition given in \S\ref{s2} that 
$\{ A^c_1\cdots A^c_N; A^o_{1} \cdots A^o_{M}\}$ is invariant
under arbitrary exchanges $A^c_i\leftrightarrow A^c_j$ and $A^o_k\leftrightarrow A^o_\ell$.
Note that there is no additional sign even for the grassmann odd vertex operators $A^o_k$.
More discussion on this can be found in comment \ref{c1} at the end of this section.

We also define
\be
[A^c_1\cdots A^c_N; A^o_{1} \cdots A^o_{M}]^c\in \wt\HH^c, \qquad
[A^c_1\cdots A^c_N; A^o_{1} \cdots A^o_{M}]^o\in\wt\HH^o\, ,
\qquad [~]_D\in \HH^c\, ,
\ee
via 
\ben\label{edefsq}
&& \langle A_0^c| c_0^-|[A^c_1\cdots A^c_N; A^o_{1} \cdots A^o_{M}]^c \rangle
 = \{ A^c_0 A^c_1\cdots A^c_N; A^o_{1} \cdots A^o_{M}\}, \quad \forall \,  |A_0^c\rangle \in \HH^c, \nonumber \\
&& \langle A_0^o|[A^c_1\cdots A^c_N; A^o_{1} \cdots A^o_{M}]^o \rangle
=  \{ A^c_1\cdots A^c_N;  A^o_0 A^o_{1} \cdots A^o_{M}\},  
\quad  \forall \,   |A_0^o\rangle \in \HH^o\,, \non\\
&& \langle \wt A^c |c_0^-| [~]_D\rangle = \{ \wt A^c\}_D\, .
\een
Here $\langle A|$ denotes the BPZ conjugate of $|A\rangle$, generated by $z\to 1/z$ for closed strings
and $z\to -1/z$ for open strings.
We shall see in \refb{edefsqd} that $[~]_D$ is related to the boundary state of the D-brane system via 
appropriate
picture changing operation and rescaling. 
It follows from \refb{edefsq}, and the fact that $\{ A^c_1\cdots A^c_N; A^o_{1} \cdots A^o_{M}\}$ is grassmann 
even, that $[A^c_1\cdots A^o_{M}]^c$ is grassmann odd and 
$[A^c_1\cdots  A^o_{M}]^o$ is grassmann even.
Similarly one can show that $| [~]_D\rangle$ is
grassmann odd.
Other useful identities involving $\{\cdots\}$, $[\cdots]$,
$\{\wt A^c\}_D$ and $[~]_D$ are:
\be\label{idensup1}
\{A_{1}^c\cdots A_{k}^c\, \GG[B_1^c\cdots B_\ell^c;C^o_1\cdots C_m^o]^c; D_{1}^o\cdots D_{n}^o\}%\nonumber\\[.3cm]
=%& =&
\{B_1^c\cdots B_\ell^c\, \GG[A_{1}^c\cdots A_{k}^c;D^o_1\cdots D_n^o]^c; C_{1}^o\cdots, C_{m}^o\}\, ,
\ee
\be\label{idensup2}
\{A_{1}^c\cdots A_{k}^c; \GG[B_1^c\cdots B_\ell^c;C^o_1\cdots C_m^o]^o D_{1}^o\cdots D_{n}^o\}=\{B_1^c\cdots B_\ell^c; 
\GG[A_{1}^c\cdots, A_{k}^c;D^o_1\cdots D_n^o]^o C_{1}^o\cdots C_{m}^o\}\, .
\ee
and
\be \label{idensup3}
\{A^c_1\cdots A^c_N[~]_D;A^o_{1} \cdots A^o_{M}\}
= \{[A^c_1\cdots A^c_N;A^o_{1} \cdots A^o_{M}]^c\}_D\, .
\ee
In \refb{idensup1}, \refb{idensup2}, $\GG$ is given by:
\ben \label{edefgg}
&& \GG|s^o\rangle =\begin{cases} |s^o\rangle \quad \hbox{if $|s^o\rangle\in \HH_{-1}$}\cr
{1\over 2} (\XX_0+\bar \XX_0) \, |s^o\rangle \quad \hbox{if $|s^o\rangle\in \HH_{-3/2}$}
\end{cases}\, , \nonumber \\ &&
\GG|s^c\rangle =\begin{cases} |s^c\rangle \quad \hbox{if $|s^c\rangle\in \HH_{-1,-1}$}\cr
\XX_0\, |s^c\rangle \quad \hbox{if $|s^c\rangle\in \HH_{-1,-3/2}$}\cr 
\bar\XX_0\, |s^c\rangle \quad \hbox{if $|s^c\rangle\in \HH_{-3/2, -1}$}\cr 
\XX_0\bar\XX_0\, |s^c\rangle \quad \hbox{if $|s^c\rangle\in \HH_{-3/2, -3/2}$}\cr 
\end{cases}\, ,
\een
with
\be \label{edefggr}
\XX_0 \equiv \ointop {dz\over z}\, \XX(z), \quad \bar\XX_0 
\equiv \ointop {d\bar z\over \bar z}\, 
\bar\XX(\bar z),
\ee
$\XX$ and $\bar\XX$ being holomorphic and anti-holomorphic PCOs. $\ointop$ includes
multiplicative factors of $\pm (2\pi i)^{-1}$.

In the following we shall also need to deal with states of wrong grassmann parity --
closed string states which are grassmann odd and open string states which are
grassmann even. To derive the relevant relations, 
we multiply each grassmann odd closed string state and grassmann even
open string state by grassmann odd $c$-numbers so that they acquire standard grassmann parities 
and
therefore obey the standard symmetry properties and other identities described above. We can now 
bring the grassmann odd  $c$-numbers to the extreme left in both sides of 
the equations keeping track of the signs
picked up during this process. 
In doing this we follow the convention that a grassmann
odd $c$-number can be passed through $\{$ and $\langle$ without any extra sign, -- the physical
origin of these rules will be described above \refb{erhs}.
Once this is done, we can remove these $c$-numbers from both sides
of the equations and derive the relevant identities. To avoid confusion, we shall use the
convention that  string states labelled by roman letters, like 
$A_i^c$, $\wt A^c$,
$A_i^o$ and $\wt A^o$,  carry the correct grassmann parity -- even for 
$A_i^c$, $\wt A^c$ and odd for $A_i^o$, $\wt A^o$. When 
we need to use states of general grassmann
parity, we shall use caligraphic letters $\AAA_i^c$, $\wt \AAA^c$,
$\AAA_i^o$ and $\wt \AAA^o$.
In this convention, in an expression like $Q_B A_i^c$, it will be understood that $A_i^c$ is grassmann even
and therefore $Q_B A_i^c$ is grassmann odd.

It can be shown that with this prescription, the generalizations of
\refb{edefsq} remain relatively simple even for states of wrong grassmann parity,
provided
we use the prescription that 
$[$ for $[\cdots]^c$ behaves as a grassmann odd object and
$[$ for $[\cdots]^o$ behaves as a
grassmann even object. In that case, 
when we replace
$A_i^c$, $\wt A^c$,
$A_i^o$ and $\wt A^o$ by $\AAA_i^c$, $\wt \AAA^c$,
$\AAA_i^o$ and $\wt \AAA^o$, 
the first and the third equations of \refb{edefsq}
remain the same, and the second equation also remains the same if
either $\AAA_0^o$ or the product $\AAA_1^c\cdots \AAA_N^c$ is grassmann even. 
In case both of these are grassmann odd,
we have an extra minus sign on the right hand side of the second equation.
To see the necessity of assigning odd grassmann parity for $[$ in $[\cdots]^c$, let us use the
generalization of the first equation of \refb{edefsq} to get
\be 
\{ \AAA_0^c \zeta \cdots\} = \langle \AAA^c_0 | c_0^- |[\zeta\cdots]^c\rangle\, ,
\ee
for a grassmann odd $c$-number $\zeta$. Now on the left hand side we can bring $\zeta$ to the extreme
left by picking up a multiplicative factor given by the grassmann parity of $\AAA^c_0$. 
On the right hand side we shall
get an additional minus sign while moving $\zeta$ through $c_0^-$. Therefore to compensate for this we need
the rule:
\be
[\zeta\cdots]^c = -\zeta[\cdots]^c\, .
\ee
There is no such factor for $[\cdots]^o$ due to the absence of $c_0$ factor in the second line of \refb{edefsq}.

Another rule we need to follow is to never move a grassmann odd variable through  $]^{c,o}$ or $\}$. To see this
consider $\{;A_1^o A_2^o A_3^o\}$ and move a grassmann odd
$c$-number $\zeta$ through this from left to right.  We may conclude that this operation will
pick up a minus sign due to three 
grassmann odd open string vertex
operators inserted in between $\{\cdots \}$. 
But this may not be the correct result, {\it e.g.} the correlation function of three grassmann odd operators on a
disc is a grassmann even number.
Furthermore, this cannot be compensated by simply assigning an
odd grassmann parity to $\}$ since the result depends on the number of boundaries. The same rule will be
followed for matrix elements like $\langle A^o| \OO |B^o\rangle$ for any operator $\OO$ acting on the open string
states.
Any grassmann odd $c$-number from inside the matrix element will be taken outside the matrix element
from the left. However 
in this case we could allow grassmann odd $c$-numbers to be taken out from the right by picking an extra minus
sign, i.e.\ by  treating the ket vacuum as grassmann odd.

We are now ready to describe the form of the action.
We introduce two sets of grassmann even closed string fields $\Psi^c\in\HH^c$ and
$\wt\Psi^c\in\wt\HH^c$ and two sets of grassmann odd open string fields
$\Psi^o\in\HH^o$ and
$\wt\Psi^o\in\wt\HH^o$.
The 1PI effective action is given by:
\ben\label{e1.21}
S_{1PI} &=& -{1\over 2\, g_s^2} \langle \wt \Psi^c | c_0^- Q_B \GG |\wt \Psi^c\rangle 
+{1\over g_s^2} \langle \wt \Psi^c | c_0^- Q_B | \Psi^c\rangle 
-{1\over 2\, g_s} \langle \wt \Psi^o | Q_B \GG |\wt\Psi^o\rangle
+ {1\over g_s} \langle \wt \Psi^o | Q_B |\Psi^o\rangle
\nonumber \\ &&
+ \{\wt \Psi^c\}_D +  \sum_{N\ge 0} \sum_{M\ge 0} {1\over N! \, M!} \, \{ (\Psi^c)^N; (\Psi^o)^{M}\}\, ,
\een
where $Q_B$ represents the closed string BRST operator on $\HH^c$ and $\wt\HH^c$ and 
open string BRST operator on $\HH^o$ and $\wt\HH^o$.
We show in \S\ref{sgauge} that the action is invariant under the gauge transformation:
\ben\label{e1.22}
|\delta \Psi^c\rangle &=& Q_B|\Lambda^c\rangle + g_s^2 \, \sum_{N,M} {1\over N!\, M!} 
\GG [\Lambda^c (\Psi^c)^N ; (\Psi^o)^M]^c +  g_s^2\, \sum_{N,M} {1\over N!\, M!} 
\GG [(\Psi^c)^N ; \Lambda^o (\Psi^o)^M]^c\, ,
\nonumber \\
|\delta \Psi^o\rangle &=& Q_B|\Lambda^o\rangle - g_s\, \sum_{N,M} {1\over N!\, M!} 
\GG [\Lambda^c (\Psi^c)^N ; (\Psi^o)^M]^o - g_s\, \sum_{N,M} {1\over N!\, M!} 
\GG [(\Psi^c)^N ; \Lambda^o (\Psi^o)^M]^o\, ,\nonumber \\
|\delta \wt\Psi^c\rangle &=& Q_B|\wt\Lambda^c\rangle + g_s^2 \, \sum_{N,M} {1\over N!\, M!} 
[\Lambda^c (\Psi^c)^N ; (\Psi^o)^M]^c +  g_s^2\, \sum_{N,M} {1\over N!\, M!} 
[(\Psi^c)^N ; \Lambda^o (\Psi^o)^M]^c\, ,
\nonumber \\
|\delta \wt\Psi^o\rangle &=& Q_B|\wt\Lambda^o\rangle - g_s\, \sum_{N,M} {1\over N!\, M!} 
[\Lambda^c (\Psi^c)^N ; (\Psi^o)^M]^o - g_s\, \sum_{N,M} {1\over N!\, M!} 
[(\Psi^c)^N ; \Lambda^o (\Psi^o)^M]^o\, , \nonumber \\
\een
where $|\Lambda^c\rangle\in\HH^c$, $|\Lambda^o\rangle\in\HH^o$, 
 $|\wt\Lambda^c\rangle\in\wt\HH^c$, $|\wt\Lambda^o\rangle\in\wt\HH^o$, 
are gauge transformation parameters. 
$\Lambda^c$ and $\wt\Lambda^c$ are grassmann odd while $\Lambda^o$ and $\wt\Lambda^o$ are
grassmann even. We also check in \S\ref{scbv} that this form of the gauge transformation is consistent with
what we obtain from the BV formalism.

Besides the various identities mentioned earlier, we need another set of identities, known as the
`main identities'\cite{9206084}, to prove gauge invariance of the action. For grassmann even $\wt A^c$, $A_i^c$ and
grassmann odd $A_i^o$, these identities take the form:
\be \label{emainex}
\{(Q_B \wt A^c)\}_D=0\, ,
\ee
and,
\ben\label{emainsup}
&& \sum_{i=1}^N \{ A^c_1\cdots A^c_{i-1} (Q_B A^c_i) A^c_{i+1}\cdots
A^c_N; A^o_{1} \cdots A^o_{M}\} \nonumber \\ &&
\hskip 1in +\sum_{j=1}^M \{ A^c_1\cdots A^c_N; A^o_{1} \cdots A^o_{j-1} (Q_B A^o_j) A^o_{j+1} \cdots A^o_M\} 
(-1)^{j-1}\nonumber \\
&=&-{1\over 2} \sum_{k=0}^N 
\sum_{\{i_1\cdots , i_k\} \subset \{1,\cdots , N\}} \sum_{\ell=0}^M \sum_{\{j_1,\cdots , j_\ell\} 
\subset \{1,\cdots, M\}} \Big(g_s^2 \, \{ A^c_{i_1} \cdots A^c_{i_k} \BB^c
; A^o_{j_1} \cdots A^o_{j_\ell}\} \nonumber \\ && \hskip 3.5 in +
g_s\, \{ A^c_{i_1} \cdots A^c_{i_k}
;  \BB^o A^o_{j_1} \cdots A^o_{j_\ell}\}\Big)\nonumber \\ &&
\hskip 1in - g_s^2 \{ [A_c^1\cdots A^c_N; A^o_1\cdots A^o_M]^c\}_D\, ,
\nonumber \\ \cr &&
\BB^c \equiv \GG [A^c_{\bar i_1} \cdots A^c_{\bar i_{N-k}}; A^o_{\bar j_1} \cdots A^o_{\bar j_{M-\ell}}]^c , 
\quad \BB^o \equiv \GG [A^c_{\bar i_1} \cdots A^c_{\bar i_{N-k}}; A^o_{\bar j_1} \cdots A^o_{\bar j_{M-\ell}}]^o\, ,
\nonumber \\ && 
\hskip -.15in 
\{i_1,\cdots , i_k\} \cup \{\bar i_1,\cdots , \bar i_{N-k}\} = \{1,\cdots, N\}, \quad
\{j_1,\cdots, j_\ell\} \cup \{\bar j_1,\cdots , \bar j_{M-\ell}\} = \{1, \cdots, M\}\, . \nonumber \\
\een
The term proportional to $\{ [A_c^1\cdots A^c_N; A^o_1\cdots A^o_M]^c\}_D$ is 
not present in bosonic open-closed
string field theory. There we include  the contribution of the disc one
point function in the definition of $\{\BB^c;\}$. As mentioned below \refb{edeffcho}, this is not an option 
for RR sector
$\BB^c$ in superstring theory due to an obstruction associated with the picture number conservation.

Note that on the left hand side of \refb{emainsup}, $Q_BA^c_i$ and $Q_BA^o_i$ are 
string states of `wrong grassmann parity', while on the right hand side $\BB^c$ and $\BB^o$ are also of
wrong grassmann parity. Therefore the corresponding objects $\{\cdots\}$ will have to be defined
by multiplying these wrong parity objects by grassmann odd $c$-number $\zeta$. 
It is useful to include this grassmann odd $c$-number $\zeta$ explicitly in the identity so that each
term in the identity has only states of correct grassmann parity. This takes the form:
\ben\label{emainsupzeta}
&& \sum_{i=1}^N \{ A^c_1\cdots A^c_{i-1} (\zeta Q_B A^c_i) A^c_{i+1}\cdots
A^c_N; A^o_{1} \cdots A^o_{M}\} \nonumber \\ &&
\hskip 1in +\sum_{j=1}^M \{ A^c_1\cdots A^c_N; A^o_{1} \cdots A^o_{j-1} (\zeta Q_B A^o_j) A^o_{j+1} \cdots A^o_M\} 
\nonumber \\
&=&-{1\over 2} \sum_{k=0}^N 
\sum_{\{i_1\cdots , i_k\} \subset \{1,\cdots , N\}} \sum_{\ell=0}^M \sum_{\{j_1,\cdots , j_\ell\} 
\subset \{1,\cdots, M\}} \Big(g_s^2 \, \{ A^c_{i_1} \cdots A^c_{i_k} B^c
; A^o_{j_1} \cdots A^o_{j_\ell}\} \nonumber \\ && \hskip 3.5 in +
g_s\, \{ A^c_{i_1} \cdots A^c_{i_k}
;  B^o A^o_{j_1} \cdots A^o_{j_\ell}\}\Big)\nonumber \\ &&
\hskip 1in - g_s^2 \{ \zeta [A_c^1\cdots A^c_N; A^o_1\cdots A^o_M]^c\}_D\, ,
\nonumber \\ \cr &&
B^c \equiv \GG \zeta [A^c_{\bar i_1} \cdots A^c_{\bar i_{N-k}}; A^o_{\bar j_1} \cdots A^o_{\bar j_{M-\ell}}]^c , 
\quad B^o \equiv \GG \zeta [A^c_{\bar i_1} \cdots A^c_{\bar i_{N-k}}; A^o_{\bar j_1} \cdots A^o_{\bar j_{M-\ell}}]^o\, ,
\nonumber \\ && 
\hskip -.15in 
\{i_1,\cdots , i_k\} \cup \{\bar i_1,\cdots , \bar i_{N-k}\} = \{1,\cdots, N\}, \quad
\{j_1,\cdots, j_\ell\} \cup \{\bar j_1,\cdots , \bar j_{M-\ell}\} = \{1, \cdots, M\}\, . \nonumber \\
\een
The $(-1)^{j-1}$ factor on the left hand side of \refb{emainsup}
will be generated when we pull $\zeta$ through the open string vertex
operators $A_i^o$ to the extreme left.

The equations of motion following from the 1PI effective action are given by
\ben \label{eeoma}
|\tilde\Psi^c\rangle\ \quad&:&\qquad Q_B\bigl(|\Psi^c\rangle-\mathcal G|\tilde\Psi^c\rangle\bigl)\ +\ g_s^2|[\ ]_D\rangle\ =\ 0\, , \\[.3cm]
\label{eeomb}
|\Psi^c\rangle\ \quad&:&\qquad Q_B|\tilde\Psi^c\rangle +g_s^2\sum_{N=1}^\infty\sum_{M=0}^\infty\f{1}{(N-1)!M!}[(\Psi^c)^{N-1};(\Psi^o)^M]^c\ =\ 0\, ,\\[.3cm]
\label{eeomc}
|\tilde\Psi^o\rangle\ \quad&:&\qquad Q_B\bigl(|\Psi^o\rangle-\mathcal G|\tilde\Psi^o\rangle\bigl)\  =\ 0\, , \\[.3cm]
\label{eeomd}
|\Psi^o\rangle\ \quad&:&\qquad Q_B|\tilde\Psi^o\rangle +g_s\sum_{N=0}^\infty\sum_{M=0}^\infty\f{1}{N!(M-1)!}[(\Psi^c)^{N};(\Psi^o)^{M-1}]^o\ =\ 0  \, .
\een
Now, multiplying the second equation by $\mathcal G$ and adding to the first equation gives
\ben\label{eeom1}
 Q_B|\Psi^c\rangle\ +\ g_s^2\sum_{N=1}^\infty\sum_{M=0}^\infty\f{1}{(N-1)!M!}\mathcal G[(\Psi^c)^{N-1};(\Psi^o)^M]^c\ +\ g_s^2|[\ ]_D\rangle\ =\ 0\, .
\een
Similarly, multiplying the 4th equation by $\mathcal G$ and adding to the 3rd equation gives
\ben\label{eeom2}
 Q_B|\Psi^o\rangle\ +\ g_s\sum_{N=0}^\infty\sum_{M=1}^\infty\f{1}{N!(M-1)!}\mathcal G[(\Psi^c)^{N};(\Psi^o)^{M-1}]^o\ =\ 0\, .
\een
Eqs.\refb{eeom1} and \refb{eeom2} give the interacting equations of motion for the physical string 
fields $\Psi^c$ and $\Psi^o$. Once we pick a solution to these equations and fix $|\Psi^c\rangle$ and
$|\Psi^o\rangle$, we can determine
$\wt\Psi^c$ and $\wt\Psi^o$ using \refb{eeomb} and \refb{eeomd} respectively. The only
freedom in obtaining solution to \refb{eeomb} is to add solutions to 
$Q_B|\wt\Psi^c\rangle=0$ to an existing solution. Similarly the only
freedom in obtaining solution to \refb{eeomd} is to add solutions to 
$Q_B|\wt\Psi^o\rangle=0$ to an existing solution. These represent  free field degrees
of freedom. Furthermore the choice of solutions to these free field equations of motion
does not affect the interacting field equations \refb{eeom1} and \refb{eeom2} since
they do not involve $\wt\Psi^c$ and $\wt\Psi^o$. Therefore the 
degrees of freedom associated with
$\wt\Psi^c$ and $\wt\Psi^o$ represent free fields that completely decouple from the
interacting part of the theory, and they have no observable signature.

Besides the results reviewed above, this paper also contains the following results:
\begin{enumerate}
\item We show in \S\ref{scbv}
that the action \refb{e1.21} satisfies the classical BV master equation
$(S,S)=0$ where $(F,G)$ denotes the  anti-bracket between two functions $F$ and
$G$ of the string field, as defined in \refb{eanticonpre}, \refb{eanticon}.
\item We construct in \S\ref{sbv} the quantum BV master action of superstring
field theory of open and closed strings. This action has the same form as \refb{e1.21}, but with  the
interaction vertices $\{\cdots\}$ replaced by slightly modified vertices $\{\hspace*{-.05in}\{\cdots\}
\hspace*{-.05in}\}$. 
Also the main identity 
satisfied by the  new vertices now has
additional terms given by the last two terms in \refb{e5.5}. 
Due to these additional terms in the vertices
the action now satisfies the quantum BV master equation \refb{eqbv}, with
$\Delta$ defined in \refb{edefDelta}.
\item In \S\ref{sun} we describe how to generalize the construction of the 1PI action and
the BV master action to unoriented open closed string field theory. The structure of the action 
remains the same but the definitions of the interaction terms change. In particular $\{A^c_1\cdots A^c_N;
A^o_1\cdots A^o_M\}$ now gets additional contribution from non-orientable surfaces, $\{\wt A^c\}_D$ gets additional
contribution from the projective sphere and there are additional normalization factors in the definitions of these
quantities.
\end{enumerate}

We end this section with a few comments.
\begin{enumerate}
\item
In most formulation of classical open string field theory an associative $*$-product 
and its generalization
known as the $A_\infty$ algebra plays a significant 
role\cite{wittensft,9503099,9705038,9707051,1201.1761,1312.2948,
1505.01659,1505.02069,
1508.00366,1510.00364,1602.02582,1602.02583}. 
This is not manifest in the formulation of the quantum action described above.
Instead what plays a central role here is the associated
Lie algebra and its infinite dimensional generalization -- the $L_\infty$ algebra. For example, if
we consider Witten's open bosonic string field theory\cite{wittensft}, 
then our definition of $[;A^o B^o]^o$ at the
tree level corresponds to $A^o*B^o - (-1)^{AB} B^o*A^o$ in the language of $*$-product.
The price we pay in giving up the $A_\infty$ structure is that we can no
longer extract color ordered amplitudes from the theory without digging into the detailed definition
of $\{\cdots\}$.\footnote{We thank Theodore 
Erler for pointing this 
out to us.} If we want to make manifest the information on color ordering, 
we need to follow a more elaborate approach
described in \cite{9705241,0410291,1109.4101}. \label{c1}

\item A special feature of the 1PI effective
action \refb{e1.21} is the linear term $\{\wt\Psi^c\}_D$ that involves
a closed string field from $\wt\HH^c$.
Physically this is related to the fact that $\Psi^c$ contains RR field strengths 
while $\wt\Psi^c$ contains the RR potential. Since D-branes carry RR charge; 
we need the RR potential, hidden inside $\wt\Psi^c$, to describe the
coupling of closed string fields to D-branes.

\item As has already been alluded to before, we can set $\{\wt A^c\}_D$ to zero if 
$\wt A^c$ belongs to the NSNS sector, provided we include this contribution into the
definition of $\{A^c;\}$. This corresponds to removing
part of $[~]_D$ that belongs to the NSNS sector
and absorbing this into the definition of $[~;~]^c$. Under this \refb{eeom1} remains unchanged.
This describes a superstring field
theory that is equivalent to the original superstring field theory.

\item Using the open-closed superstring field theory one can construct gauge invariant 1PI effective
actions for theories that are apparently anomalous. For example if we consider type IIB string theory with
certain number of space-filling D9-branes, its spectrum will contain,
besides the usual closed string fields, additional chiral fermions from the gauge
supermultiplet on the D9-branes. This theory is  known to suffer from gravitational anomaly\cite{alvarez-witten}.
However there is no difficulty in writing down a gauge invariant 1PI effective action for this theory. In the latter
description, the inconsistency shows up due to the presence of a term in the action
that is linear in the RR 10-form field,
encoded in the $\{\wt\Psi^c\}_D$ term in \refb{e1.21}. Due to the presence of this term, the theory does not have
a vacuum solution to the equations of motion\cite{9711087}.  
However in some cases we may be able to cancel the effect
of the $\{\wt\Psi^c\}_D$ term, leading to the last term on the left hand side of \refb{eeom1}, 
by switching on other background fields
contributing to the second term on the left hand side of \refb{eeom1}. Examples of this kind can be found
in compactification of type IIB string theory on Calabi-Yau manifolds where the space-filling D3-brane charge
can be cancelled by flux of 3-form fields along the internal 3-cycles of the Calabi-Yau manifold.

\item The action \refb{e1.21} contains insertion of $c_0^-$ in several places -- in the kinetic term as well as in the
definition of $\{\wt\Psi^c\}_D$ given in \refb{e1.5a}. This can be traced to the presence of conformal Killing vectors on
the associated Riemann surfaces -- a sphere with two punctures and a line integral of the BRST current, 
and a disk with one bulk
puncture. We can avoid this 
by including $c_0^-$ in the definition of $\wt\Psi^c$, declaring $\wt\HH^c$ to be 
the subspace of states 
annihilated by $c_0^-$ instead of $b_0^-$. This will introduce a $b_0^-$ in the $\wt\Psi^c$-$\wt\Psi^c$
kinetic term, but this is a more natural operator since the anti-commutator of $Q_B$ with $b_0^-$ generates
$L_0^-$. It has in fact been argued in \cite{yuji} that in this formalism we can take
$\Psi^c$ and $\wt\Psi^c$ to be unconstrained elements in the Hilbert space of
closed string states except for the restriction on the picture numbers. 
This introduces some additional free field degrees of freedom which
decouple from the interacting part of the theory.

\item As has been mentioned already, \refb{eeom1} and \refb{eeom2} 
can be  regarded as the
equations of motion of the physical string fields $\Psi^c$ and $\Psi^o$. 
One important question is: 
given a solution to these equations, can we always find a solution to the 
equations of motion \refb{eeomb} and \refb{eeomd} for the
additional fields $\wt\Psi^c$ and
$\wt\Psi^o$? If the answer is in the negative, this may impose further 
constraints on the physical fields $\Psi^c$ and $\Psi^o$. 
Now, using \refb{emainsup} it is straightforward to show that once
\refb{eeom1} and \refb{eeom2} are satisfied, the second terms in \refb{eeomb} and
\refb{eeomd} are
BRST invariant. 
Therefore the question reduces to whether they describe 
non-trivial elements of the BRST cohomology, since as long as they are BRST
trivial one can always find $\wt\Psi^c$ and $\wt\Psi^o$ 
satisfying \refb{eeomb} and \refb{eeomd}. 
This can be studied
separately in the zero momentum sector and the non-zero momentum sector.
We shall analyze
this question for \refb{eeomb}, -- the analysis of \refb{eeomd} will be similar and
in fact simpler. 
In the sector carrying non-zero momentum along the non-compact
space-time directions, the contribution from $[~]_D$ can be expressed as
$Q_B b_0^+ (L_0^+)^{-1}[~]_D$,
and therefore the second term in \refb{eeom1}, being equal to
$-Q_B|\Psi^c\rangle - Q_B b_0^+ (L_0^+)^{-1}[~]_D$,
 is BRST trivial. It follows from the
analysis of \cite{9711087} that the second term in \refb{eeomb} is also BRST trivial, --
one can use the inverse picture changing operator introduced in \cite{9711087} to
map elements of the BRST cohomology from $\HH^c$ to $\wt\HH^c$ in the 
non-zero momentum sector.  On the other hand, it was shown in \cite{9711087} that
in $\wt\HH^c$, the BRST cohomology in the zero momentum sector is trivial.\footnote{For
this one needs to use $x_0$ cohomology where one allows polynomials
in the zero modes of the 
non-compact space-time coordinate fields to appear in the state.} Therefore the
BRST invariance of the second term in \refb{eeomb} implies  BRST exactness
of this term. This, in turn, implies that \refb{eeomb} always has a solution when 
\refb{eeom1} and \refb{eeom2} are satisfied. A similar conclusion follows for \refb{eeomd}.
Therefore \refb{eeomb} and \refb{eeomd} do not impose any additional constraint on $\Psi^c$
and $\Psi^o$ besides the ones implied by \refb{eeom1} and \refb{eeom2}.

\item Once a consistent superstring field theory for open and closed strings has been formulated, it can be used
to systematically study various aspects of string theory that are not easily amenable to the standard 
world-sheet approach. This includes for example the study of mass renormalization or vacuum shift\cite{1703.06410}, 
or studying superstring theory in RR background\cite{1811.00032}.

\item The superstring field theory of open and closed strings constructed here does not suffer from any 
ultra-violet divergence. However this theory suffers from all the usual infra-red divergences that a 
quantum field theory suffers. These have physical origin and need to be dealt with as in a quantum
field theory.

\item Superstring field theory action 
that we write down can be formulated around any background associated with a
superconformally invariant world-sheet theory in the NSR formalism.
Even for a given background the theory is not unique -- it depends on the choice
of local coordinates at the punctures and the choice of PCO locations that we 
have to make in defining the interaction terms of the action. For superstring field 
theories of closed strings it is known that apparently different string field theories, that
one gets by making different choices, are all related by field redefinition. We expect
a similar result to hold for the theory described here, but we have not attempted to
give a complete proof. Another feature of the theory that we would like to prove is
background independence -- if we have different backgrounds  related by marginal
deformation of the world-sheet theory, then the superstring field theories formulated
around these different backgrounds should also be related by field redefinition. This
analysis is expected to be more complicated than the corresponding analysis for closed 
superstring field theory, since under marginal deformation of the bulk world-sheet
theory we also need to deform the D-brane system appropriately\cite{9705241}. 
We hope to
return to these problems in the future.

\end{enumerate}

\sectiono{Construction of 1PI vertices and their properties} \label{s2}

We shall now describe the construction of the 1PI vertices $\{ A^c_1\cdots A^c_N; A^o_{1} \cdots A^o_{M}\}$
and $\{\wt A^c\}_D$
satisfying the various identities described in \S\ref{s1}. Since the construction proceeds more or less in the same
way as in the case of closed superstring field theory reviewed in \cite{1703.06410}, 
we shall only emphasize the 
differences.

We begin by describing our convention for correlation functions of
open string vertex operators.
If there are $N$ D-branes, not necessarily of the same kind, then a general open string state is described by
an $N\times N$ matrix, with the $i$-$j$ matrix element describing the state of an open string whose left end is on the $i$-th D-brane and the right end is 
on the $j$-th D-brane. A correlation function of open and closed string vertex operators  on a general
Riemann surface with multiple boundaries include taking traces over the $N\times N$ Chan-Paton matrices
on each boundary of the Riemann surface. These traces will not be written explicitly, 
but will be understood as part of the definition of the correlation function. We shall follow the standard
convention of referring to the locations of the vertex operators on the Riemann surface as punctures,
with closed string vertex operators inserted at the bulk punctures and the open string vertex operators
inserted at the boundary punctures.

We shall first define 
$\{\wt A^c\}_D$. We take this to be the 
one point function of $g_s^{-1} c_0^- \wh\GG \wt A^c$ on the unit disc
$|z|\le 1$, with the vertex operator inserted at the center of the disc using local
coordinate $z\, e^\beta$. Here $\beta$ is some positive number and\footnote{Since there is no insertion
on the boundary, $\XX_0$ and $\bar \XX_0$ could be evaluated by taking their defining integration contours on the
boundary. Since on the boundary the holomorphic and anti-holomorphic PCOs are equal, $\XX_0$ and $\bar\XX_0$
are identical. Therefore we could replace $\wh\GG$ by $\XX_0$ or $\bar \XX_0$.}
\be\label{edefgghat}
 \wh\GG \equiv \begin{cases}
\hbox{1 on $\HH_{-1,-1}$}\cr
{1\over 2} (\XX_0+\bar \XX_0) \, \hbox{on $\HH_{-3/2,-3/2}$}
\end{cases}\, ,
\ee
where $\XX_0$, $\bar \XX_0$ are zero modes of the
PCOs as defined in \refb{edefggr}.
The $c_0^-$ factor is needed due to the presence of conformal Killing vector on the
disc, which makes one point function of any operator vanish if the operator is annihilated by
$b_0^-$. The $\wh\GG$ factor is needed for picture number conservation. Since
$\wt A^c$ has picture number $-2$ in the NSNS sector and $-3$ in the RR sector,
$\wh\GG A^c$ has picture number $-2$ in both sectors. This is the
correct picture number for getting a non-vanishing one point function on the disc.
The identity \refb{emainex} can now be proved by deforming the integration contour defining $Q_B$ to the boundary
of the disc. On the boundary the holomorphic and the anti-holomorphic components of $Q_B$ cancel each other.
There is no additional contribution from having to pass $Q_B$ through $\wh\GG$ since $[Q_B,\wh\GG]$
vanishes. The term containing $\{Q_B, c_0^-\}$ vanishes since $\{Q_B, c_0^-\}$ does not contain a $c_0^-$ and
therefore integration over the zero mode $c_0^-$ associated with the conformal Killing vector vanishes. 

Given this definition of $\{\wt A^c\}_D$,  we can define $[~]_D$
via \refb{edefsq}.
We can also express $[~]_D$ in terms of the boundary state\cite{boundary1,boundary2,boundary3,boundary4}
of the D-brane system as follows.
Let $|B\rangle$ be the boundary state of the D-brane system under
consideration so that 
$\langle \phi^c|c_0^-|B\rangle$ describes the one point function
of the closed string vertex operator $c_0^-\phi^c$,
inserted at the center of the unit disc $|z|\le 1$ in the 
$z$ coordinate system. This gives
\be\label{e1.5a}
\{\wt A^c\}_D = g_s^{-1}\, \langle\wt A^c|c_0^-\, \wh\GG \, e^{-\beta(L_0+\bar L_0)} |B\rangle\, .
\ee
Since one point function on the disc is non-zero only for vertex operators of
picture number $-2$, 
$\phi^c$ must have picture number $-2$ for getting non-vanishing $\langle\phi^c|c_0^-|B\rangle$.
On the other hand, 
$\langle \phi^c|c_0^-|B\rangle$, being related to a sphere correlation function
of $\phi^c$ and $B$,
is non-zero only when the total picture number
of $\phi^c$ and $B$ add up to $-4$. Therefore $|B\rangle$ has picture number $-2$. Comparing the
last equation of \refb{edefsq}
and \refb{e1.5a}, we can express
$[]_D\in \HH^c$ as
\be \label{edefsqd}
[]_D = g_s^{-1} {\bf P} \, \wh\GG \, e^{-\beta (L_0+\bar L_0)} 
|B\rangle\, ,
\ee
where $\bf P$ denotes projection operator into $\HH^c$. Its role is to pick the $(-1,-1)$
and $(-1/2,-1/2)$ components of $\wh \GG |B\rangle$, which otherwise has states in 
$\HH_{m,n}$ for all $m,n$ with $m+n=-2$ in the NSNS sector and $m+n=-1$ in the RR sector.

The construction of  $\{\cdots\}$ proceeds 
as in the case of closed superstring field theory -- therefore
we shall be brief, emphasizing only the new aspects of this construction.
We denote by $\MM_{g,b,m_c,n_c,p_c,q_c,m_o,n_o}$ the moduli space of Riemann surfaces with genus 
$g$, $b$ 
boundaries, $m_c$ NSNS punctures, $n_c$ NSR punctures, $p_c$ RNS punctures, $q_c$ RR punctures, 
$m_o$ NS-sector punctures on the boundary and $n_o$ R-sector punctures on the boundary, with the
understanding that the 
integration over $\MM_{g,b,m_c,n_c,p_c,q_c,m_o,n_o}$ includes sum over spin structures.
It will also be understood that the Ramond punctures carry picture number $-1/2$, i.e. when we insert a vertex 
operator at the puncture, it belongs to $\HH^c$ or $\HH^o$.
We denote by
$\wt\PP_{g,b,m_c,n_c,p_c,q_c,m_o,n_o}$ a fiber bundle over this moduli space, with the fiber directions 
specifying the choice of local coordinates at the punctures and also the locations of the 
\be
\wp\equiv 4g + 2b -4+ 2m_c+3(n_c+p_c)/2+q_c+m_o + n_o/2\, ,
\ee 
PCOs. Note that for Riemann surfaces with boundary, the 
picture number in the holomorphic and anti-holomorphic sectors are not separately conserved, but
only the total picture number is conserved.
For this reason, we have specified only the total picture number. Only for
$b=0$, we have to have 
$2g-2+m_c+n_c+(p_c+q_c)/2$ anti-holomorphic
PCOs and $2g-2+m_c+p_c+(n_c+q_c)/2$ holomorphic PCOs. Therefore for $b\ne 0$ the fiber has
$\wp+1$ different branches, with the $r$-th branch having $r$ holomorphic PCOs and $(\wp-r)$ 
anti-holomorphic PCOs. 

In order to simplify notation we shall define the formal sum
\be
\wt\PP_{g,b,N,M} \equiv \sum_{m_c,n_c,p_c,q_c,m_o,n_o\atop m_c+n_c+p_c+q_c=N,
m_o+n_o=M} \wt\PP_{g,b,m_c,n_c,p_c,q_c,m_o,n_o}\, ,
\ee
and similarly $\MM_{g,b,N,M}$. Following the same procedure 
reviewed in \cite{1703.06410} for closed string theory, given a set of closed string states $A^c_1,\cdots, A^c_N$ and
open string states $A^o_1,\cdots, A^o_M$, we can construct a $p$-form 
$\Omega^{g,b,N,M}_p(A^c_1,\cdots, A^c_N;A^o_1,\cdots, A^o_M)$ on
$\wt\PP_{g,b,N,M}$  for any positive integer $p$,
in terms of appropriate correlation functions of $b$-ghosts,
PCOs and of vertex
operators $\{A^c_i\}$, $\{A^o_i\}$ on the corresponding Riemann surfaces. In this
construction
the local coordinates used in the insertion of off-shell vertex operators and  the
locations of the PCOs are determined by the point in 
$\wt\PP_{g,b,N,M}$ where we compute the $p$-form. The specific form of the ghost insertions is
determined by the tangent vectors of $\wt\PP_{g,b,N,M}$ with which we contract $\Omega^{g,b,N,M}_p$,
i.e.\ the particular components of $\Omega^{g,b,N,M}_p$ that we want to compute.
The sign rules of \S\ref{s1}, that tells us that we can take a grassmann odd $c$-number
out of $\{\cdots\}$
from the left, implicitly assumes that in computing the correlation functions that define
$\Omega^{g,b,N,M}_p$, we insert the vertex operators for the external states first in the
order they appear inside $\{\cdots\}$ and then insert all the ghosts and PCOs. Similarly the sign rule
that we can move a grassmann odd $c$-number through $\langle$ without extra sign corresponds to treating
the bra vacuum $\langle 0|$ as grassmann even.
$\Omega^{g,b,N,M}_p$
satisfies the useful property:
\ben \label{erhs}
&&\sum_{i=1}^N\Omega^{g,b,N,M}_p(A^c_1,\cdots, A^c_{i-1}, Q_B A^c_i, A^c_{i+1},\cdots A^c_N;A^o_1,\cdots, A^o_M)\non\\
&& +\sum_{j=1}^M (-1)^{j-1}
\Omega^{g,b,N,M}_p(A^c_1,\cdots, A^c_N;A^o_1,\cdots, A^o_{j-1},
Q_B A^o_j, A^o_{j+1},\cdots, A^o_M)\non\\
&=& \kappa(g,b,N,M,p) \, d\Omega^{g,b,N,M}_{p-1}
(A^c_1,\cdots, A^c_N;A^o_1,\cdots, A^o_M)\, ,
\een
where $\kappa$ is an appropriate sign factor about which we shall say more later.
The identity is derived by deforming the integration contour used in defining $Q_B$ 
away from the vertex operators and making it act on the ghosts / PCO insertions. This 
generates insertion of stress tensor in the correlation function 
which in turn has the interpretation of an exterior
derivative acting on $\Omega_{p-1}$.
The phase $\kappa$ could in principle differ from the corresponding result in the
closed string case from having to move odd operators through open string vertex operators.
It can be determined by careful analysis as in \cite{9705241}, but we shall
extract the relevant information using an indirect approach to be described later. 
A special role will be
played by $\Omega^{g,b,N,M}_{6g-6+3b+2N+M}$
since the dimension of the moduli space $\MM_{g,b,N,M}$ is given by
$6g-6+3b+2N+M$.

As in the case of closed (super-)string field theory reviewed in \cite{1703.06410}, 
we introduce the notion of a generalized section of $\wt\PP_{g,b,N,M}$
by extending the
notion of a section. A generalized section can be a formal weighted 
average of many sections -- with the understanding
that integral over such a generalized section will be given by the  weighted average of the integral over the
corresponding sections. Unlike a regular section,
a generalized section may also contain vertical segments across which the
PCO locations jump discontinuously. The integral of 
$\Omega^{g,b,N,M}_{6g-6+3b+2N+M}$
over such vertical segments will have to be defined by
adding to the integral over the continuous part of the section some correction terms described in \cite{1408.0571,1504.00609}.

The ability to include vertical segments in the generalized section plays a crucial role in open-closed
string field theory -- it allows us to choose generalized sections that
can jump between the different branches of $\wt\PP_{g,b,N,M}$ mentioned earlier.
This is done
by moving one or more
PCOs from the bulk to a boundary across a vertical 
segment\cite{1408.0571,1504.00609}. 
Since on the boundary the holomorphic and anti-holomorphic 
PCOs are identical, we can replace holomorphic  PCOs by anti-holomorphic PCOs (or vice versa)
and then move them to
the desired positions in the bulk 
across another vertical segment. Such jumps may be necessary in order to ensure that
near various boundaries of the moduli space, the generalized section factorizes correctly into the direct
product of the sections on the component Riemann surfaces to which the original Riemann surface degenerates.

For defining $\{\cdots\}$ we also need the notion of a section segment where we remove from
the base $\MM_{g,b,N,M}$ certain codimension zero 
regions
and then erect a generalized section on this truncated space. 
Even though we drop the word generalized for brevity, it should be understood that the section segments
we shall be working with refer to generalized section segments, allowing us to take weighted
averages and vertical segments.
Each such section segment represents a 
family of punctured Riemann surfaces equipped with a choice of local coordinates at the punctures
and the choice of PCO locations.
We shall now define two operations on section segments that will be important for us:
\begin{enumerate}
\item {\bf Sewing:}
Let us take a pair of section segments, one in $\wt\PP_{g,b,N,M}$ and the other in
$\wt\PP_{g',b',N',M'}$.
Each represents  a family of Riemann surfaces equipped with
choice of local coordinates at the punctures and PCO locations. We can now
construct a new section segment by sewing
them at a pair of punctures -- one from  each section segment. 
If the sewing is done at a bulk puncture, then this means that we take a Riemann surface from one family
and sew one of its punctures to a puncture on the 
Riemann surface from the other family by making the identification
\be\label{esew1}
w_1 w_2 = e^{-s-i\theta}, \quad 0\le s\le \infty, \quad 0\le\theta<2\pi\, ,
\ee
where $w_1$ and $w_2$ are the local coordinates at the respective punctures.
For sewing at a pair of boundary punctures the analog of \refb{esew1} takes the form
\be\label{esew2}
w_1 w_2 = -e^{-s}\, .
\ee
In both cases,
we also insert a factor of $\GG$ defined in \refb{edefgg} around the origin of the 
$w_1$ (or the $w_2$) coordinate system. Doing this operation for each element of 
the first section segment and each element of the 
second section segment, we generate a 
family of new Riemann surfaces equipped with local coordinates at the punctures and choice of PCO
locations, producing a new section segment. 
When the sewing is done at a pair of bulk punctures via \refb{esew1}, 
the resulting family of Riemann surfaces gives a section segment of $\wt\PP_{g+g',b+b',N+N'-2,M+M'}$.
On the other hand for sewing at a pair of boundary punctures via \refb{esew2}, we get
a section segment of $\wt\PP_{g+g',b+b'-1,N+N',M+M'-2}$.

\item {\bf Hole creation:}
Another way of producing a new section segment from a given one is to
sew a Riemann surface belonging to a section segment to a disc $|z|\le 1$
with one bulk puncture,
via the relation 
\be \label{esew3}
w_1 w_2 = e^{-s}\, ,
\ee 
where $w_1$ is the local coordinate at the bulk puncture on the Riemann surface that is
being sewed and $w_2$ is the local coordinate at the bulk puncture at the center of the 
disc. We also need to insert a factor of $\wh\GG$ defined in \refb{edefsqd} around the
origin of the $w_1$ or $w_2$ coordinate system.
Note the absence of
the phase $e^{-i\theta}$ even though we sew two closed string punctures -- this is
related to the presence of a conformal Killing vector on the one punctured disc.
We shall take $w_2$ to be related to the standard coordinate $z$ on the unit 
disc by $w_2=e^{\beta} z$, where $\beta$ is the positive constant that appeared in the definition
of $\{\wt A^c\}_D$. In that case, hole
creation is equivalent to inserting at a 
bulk puncture on the Riemann surface the state (see \refb{eww3})
\be\label{esew3new}
-g_s^{2}\, e^{-s(L_0+\bar L_0)}  (b_0+\bar b_0) |[]_D\rangle
= - g_s\,  e^{-(s+\beta)(L_0+\bar L_0)} \, {\bf P}\, \wh \GG \, 
(b_0+\bar b_0) |B\rangle\, , \quad 0\le s\le \infty\, .
\ee
The parameter $s$ labels the extra modulus that
appears when we replace a closed string puncture by a boundary. If the original section segment
belonged to $\wt\PP_{g,b,N,M}$, then the new section segment obtained by hole creation belongs to
$\wt\PP_{g,b+1,N-1,M}$.

In bosonic open-closed string field theory the hole creation need not be described as a separate operation,
-- it can be included in the sewing of a Riemann surface with punctures to the disc with one bulk puncture.
In open-closed superstring field theory the disc with one bulk puncture requires special treatment since picture number
conservation makes the disc one point function of vertex operators in $\HH^c$  vanish and we need to pick
vertex  operators from $\wt\HH^c$. 
\end{enumerate}

The definition of $\{\cdots\}$ 
requires choice of section segments
$\RR_{g,b,N,M}$ of 
$\wt\PP_{g,b,N,M}$
satisfying certain properties:
\begin{enumerate}
\item The projection of $\RR_{g,b,N,M}$ on the 
base $\MM_{g,b,N,M}$ must not contain any separating type degeneration.
\item $\RR_{g,b,N,M}$ must be symmetric under the exchange of
punctures, separately for closed strings and for open strings.
\item 
Given a set of section segments $\{\RR_{g,b,N,M}\}$, we can generate new section segments 
from them by
repeated application of sewing and hole creation.
The demand we make on $\RR_{g,b,N,M}$ is that the formal
sum of all of these section segments produces a full generalized section whose projection to 
the base covers
the full moduli space $\MM_{g,b,N,M}$.
\end{enumerate}

As in the case of closed superstring field theory described in \cite{1703.06410}, 
we can systematically construct the
section segments $\RR_{g,b,N,M}$'s satisfying the above requirements as follows. 
Since the sewing and hole creation operation always increase the dimension of the section
segment due to appearance of new parameters $s$ or $\theta$, we shall set up a recursive procedure
for constructing $\RR_{g,b,N,M}$ with the dimension of $\RR_{g,b,N,M}$ as the recursion parameter. 
We begin by making
choices of $\RR_{0,0,3,0}$,  $\RR_{0,1,0,3}$ and $\RR_{0,1,1,1}$.
In each of these cases the moduli space is 0-dimensional, so the only choice we have to make is the choice
of local coordinates and PCO locations.\footnote{Note that $\RR_{0,0,n,0}$ and 
$\RR_{0,1,0,n}$ are empty for
$n\le 2$.}
We choose the local coordinates $w_i$ so that they are related to some
natural local coordinate $z$ on these Riemann surfaces by large scaling: $w_i=e^{\beta_i} \, z$ 
with large $|\beta_i|$, so
that $|w_i|=1$ describes a small circle around the puncture in the natural coordinates. This is known as adding
long stubs to the string vertices\cite{9301097}. 
Similarly we choose  the parameter $\beta$ appearing in the definition of
$\{\wt A^c\}_D$ in \refb{edefsqd} to be large. 
In order to make $\RR_{g,b,N,M}$ symmetric under the exchange of punctures, we may
need to average over different choices of local coordinates and / or PCO locations --
this is allowed since we only require $\RR_{g,b,N,M}$ to be a generalized section
segment.
We now focus on $\RR_{g,b,N,M}$'s whose expected dimension
is 1, {\it e.g.} $\RR_{0,1,0,4}$, $\RR_{0,1,1,2}$, $\RR_{0,1,2,0}$ and $\RR_{0,2,0,1}$.
For any such $g,b,N,M$, we first determine all section segments obtained by sewing or 
hole creation operation
involving zero dimensional $\RR_{g',b',N',M'}$. 
As long as the zero dimensional $\RR_{g',b',N',M'}$ have been constructed by adding long stubs,
sewing and hole creation of the corresponding section segments will generate one dimensional
section segments of $\wt\PP_{g,b,N,M}$, 
whose projection on the base covers only small regions in $\MM_{g,b,N,M}$ around separating type
degenerations, leaving behind large gaps.
We now choose one dimensional
$\RR_{g,b,N,M}$ to `fill the gap' so that together we have a complete generalized section of
$\wt\PP_{g,b,N,M}$. There is clearly a lot of freedom since the section segments generated by
sewing
or hole creation of
zero dimensional $\RR_{g',b',M',N'}$ only fix the boundaries of one dimensional
$\RR_{g,b,N,M}$, by requiring them to
match the $s=0$ boundaries of the sewing operation \refb{esew2} or hole creation 
operation \refb{esew3}. In the interior we can choose local coordinates / PCO locations arbitrarily,
subject to the restriction that 
the local coordinates 
should carry 
long stubs, and the  PCOs must avoid spurious poles\cite{verlinde} by finite
margin\cite{1703.06410}. We now repeat the process, generating all 
the two dimensional section segments by sewing 
and hole creation of the section segments of lower dimensional $\RR_{g',b',N',M'}$'s and then choose the
two dimensional $\RR_{g,b,N,M}$'s by filing the gap. This procedure can be repeated to generate all the
$\RR_{g,b,N,M}$'s.

Note that allowing $\RR_{g,b,N,M}$ to be generalized section segments is crucial for
this construction. We have already mentioned that making it symmetric under the
exchange of punctures may require averaging over different choices of
local coordinates and/or PCO locations. Furthermore the boundaries of $\RR_{g,b,N,M}$,
fixed by sewing or hole creation in lower dimensional $\RR_{g',b,M',N'}$'s, are often
generalized sections since they often have insertion of 
the operator $\GG$ or $\wh\GG$ involving 
average of PCO insertions on a circle.

For the purpose of our analysis we shall not need the explicit form of $\RR_{g,b,N,M}$'s.
Explicit construction of such $\RR_{g,b,N,M}$'s can be done using minimal area metric\cite{9206084}
or hyperbolic
metric\cite{1703.10563,1706.07366,1708.04977}, but any other 
choice satisfying the above requirements will also be acceptable for
our construction.
We now define:
\ben \label{edefcurly}
&& \{ A^c_1\cdots A^c_N; A^o_{1} \cdots A^o_{M}\} \\
&\equiv& \sum_{g,b\ge 0\atop (g,b)\ne (0,1) \, \hbox{\tiny
for}\, (N,M)=(1,0)} \, (g_s)^{2g-2+b}
\int_{\RR_{g,b,N,M}} \, \Omega^{g,b,N,M}_{6g-6+3b+2N+M}
(A^c_1,\cdots, A^c_N;A^o_1,\cdots, A^o_M)\, . \non
\een

The proof of the main identity \refb{emainsup} can now be given as follows. It is clear from
\refb{edefcurly} that the left hand side of the main identity is given by 
\be
\sum_{g,b\ge 0\atop (g,b)\ne (0,1) \, \hbox{\tiny
for}\, (N,M)=(1,0)} \, (g_s)^{2g-2+b}
\int_{\RR_{g,b,N,M}} \,  \hbox{lhs of \refb{erhs} with 
$p=6g-6+3b+2N+M$}\, .
\ee
We can now use \refb{erhs} to express this as
\be\label{e3.10}
\sum_{g,b\ge 0\atop (g,b)\ne (0,1) \, \hbox{\tiny
for}\, (N,M)=(1,0)} \, (g_s)^{2g-2+b}
\int_{\p\RR_{g,b,N,M}} \, \kappa_{g,b,N,M} \, \Omega^{g,b,N,M}_{6g-7+3b+2N+M}
(A^c_1,\cdots, A^c_N;A^o_1,\cdots, A^o_M)\, ,
\ee
where $\kappa_{g,b,N,M}\equiv \kappa(g,b,N,M,6g-6+3b+2N+M)$. From the definition of
$\RR_{g,b,N,M}$ given above, it follows that
$\p\RR_{g,b,N,M}$ must match onto the $s=0$ boundary of one of the three sewing
operations \refb{esew1}, \refb{esew2}, \refb{esew3} 
acting on (a pair of) $\RR_{g',b',N',M'}$. The standard arguments in
conformal field theory now show that up to signs, the $s=0$ boundary of the
operation \refb{esew1} produces the term involving $\BB^c$ on the right hand side of
\refb{emainsup},
the $s=0$ boundary of the
operation \refb{esew2} produces the term involving $\BB^o$ 
on the right hand side of
\refb{emainsup}, and the $s=0$ boundary of the
operation \refb{esew3} produces the term involving $\{[\cdots]\}_D$ 
on the right hand side of
\refb{emainsup}. The signs can be determined by careful analysis as in
\cite{9705241} since there is no essential difference between 
bosonic and superstring theories here, the open string fields being grassmann
odd in both cases. However we shall determine the signs by an indirect
argument. We outline below the general strategy, leaving the detailed analysis to
appendix \ref{sa}.

First of all, the fact that using \refb{emainsup} we can prove
gauge invariance of the action, as shown in \S\ref{sgauge}, 
provides an indirect evidence for the correctness of the
signs of the terms on the right hand side of \refb{emainsup}. 
We can provide a more direct argument for these signs using 
the factorization property of string
amplitudes as follows. Factorization property
tells us that near the separating type degeneration, the integrand
of a string amplitude breaks into a sum of  products of the integrands associated with the 
component Riemann surfaces into which the original Riemann surface degenerates.
This property is needed to ensure that the contribution to the amplitude from the 
region near a separating type degeneration has the interpretation
of a pair of string amplitudes connected by a propagator. This can be used to fix
any phase ambiguity in the integrand of a string amplitude by relating it to the 
product of the phases of amplitudes at lower genus / with lower number of punctures. We
shall show in appendix \ref{sa}  that if the phases of the amplitudes are fixed this way, then \refb{e3.10}
gives precisely the 
right hand side of  the main identity \refb{emainsup} without any extra sign.

Other properties of $\{\cdots\}$ and $[\cdots]^{c,o}$ described in \S\ref{s1} can be
proved easily.  For example, symmetry of $\{ A^c_1\cdots A^c_N; A^o_{1} \cdots A^o_{M}\}$ 
under arbitrary exchanges $A^c_i\leftrightarrow A^c_j$ and $A^o_k\leftrightarrow A^o_\ell$ follows
immediately from the symmetry of $\RR_{g,b,N,M}$ under permutation of bulk punctures and of
boundary punctures.
Put another way,  the definition
of $\{ A^c_1\cdots A^c_N; A^o_{1} \cdots A^o_{M}\}$ involves explicit symmetrization under
$A^c_i\leftrightarrow A^c_j$ and $A^o_k\leftrightarrow A^o_\ell$.
 The fact that $\{ A^c_1\cdots A^c_N; A^o_{1} \cdots A^o_{M}\}$ and
$\{\tilde A^c\}_D$ are grassmann even 
can be proved iteratively
by starting with an amplitude with no external states (or, for low genus, amplitudes with
minimal number of external states needed for removing the conformal Killing vectors) and then
noting that the addition of a grassmann even closed string state is accompanied by two insertions
of $b$-ghosts in the correlator, while the addition of a grassmann odd open string is accompanied
by one insertion of $b$-ghost. Therefore the grassmann parity of the correlator remains unchanged
under these operations. This is also consistent with
the fact that in order to get a non-zero result for an
amplitude, the total number of grassmann odd component fields (coming from space-time
ghosts, fermions etc.) must be even.
The identities \refb{idensup1}-\refb{idensup3} can be proved using \refb{edefsq}. 
For example we can express the left hand side of \refb{idensup1} as:
\ben\label{idensup1proof}
&& \{A_{1}^c\cdots A_{k}^c\, \GG[B_1^c\cdots B_\ell^c;C^o_1\cdots C_m^o]^c; D_{1}^o\cdots D_{n}^o\}
\non\\ &=& 
\langle \GG[B_1^c\cdots B_\ell^c;C^o_1\cdots C_m^o]^c| c_0^- |[A_{1}^c\cdots A_{k}^c;D_{1}^o\cdots D_{n}^o]^c
\rangle \non\\
&=& \langle [B_1^c\cdots B_\ell^c;C^o_1\cdots C_m^o]^c| c_0^-\,  
\GG |[A_{1}^c
\cdots A_{k}^c;D_{1}^o\cdots D_{n}^o]^c\rangle\non\\
&=& \langle \GG [A_{1}^c
\cdots A_{k}^c;D_{1}^o\cdots D_{n}^o]^c| c_0^-\,  
|[B_1^c\cdots B_\ell^c;C^o_1\cdots C_m^o]^c\rangle\non\\
&=&\{B_1^c\cdots B_\ell^c\, \GG[A_{1}^c\cdots A_{k}^c;D_{1}^o\cdots D_{n}^o]^c; C^o_1\cdots C_m^o\}
\, ,
\een
where in the second step we have used 
the fact that the both $[\cdots]^c$ in \refb{idensup1proof} are grassmann odd.
This establishes the symmetry property
\refb{idensup1}. The other identities \refb{idensup2}, \refb{idensup3} can be proven in a 
similar manner.

\sectiono{Gauge invariance of the 1PI action} \label{sgauge}

In this section we shall prove the invariance of the 1PI action 
given in \refb{e1.21} under the gauge transformation
\refb{e1.22}. The corresponding result for bosonic string field theory of open and closed
strings follows
from this by setting $\wt \Psi=\Psi$, $\wt \Lambda =\Lambda$ and $\GG=1$.

Under the gauge transformation \refb{e1.22}, the variation in the action 
\refb{e1.21} is given by
\ben
\delta S_{1PI} &=&   -\f{1}{g_s^2}\langle \wt\Psi^c|c_0^-Q_B\mathcal G|\delta\wt\Psi^c \rangle +\ \f{1}{g_s^2}\langle \delta\wt\Psi^c|c_0^-Q_B|\Psi^c \rangle \ +\ \f{1}{g_s^2}\langle \wt\Psi^c|c_0^-Q_B|\delta\Psi^c \rangle\non\\
&&-\ \f{1}{g_s}\langle \wt\Psi^o|Q_B\mathcal G|\delta\wt\Psi^o \rangle\ +\ \f{1}{g_s}\langle \delta\wt\Psi^o|Q_B|\Psi^o \rangle\ +\ \f{1}{g_s}\langle \wt\Psi^o|Q_B|\delta\Psi^o \rangle\non\\[.2cm]
&&\hskip -.5in +\, \{\delta\wt\Psi^c\}_D +\sum_{N= 1}^\infty\sum_{M= 0}^\infty\f{N}{N!M!}\{(\Psi^c)^{N-1}\delta\Psi^c;(\Psi^o)^M\} \ +\sum_{N= 0}^\infty\sum_{M= 1}^\infty\f{M}{N!M!}\{(\Psi^c)^N;\delta\Psi^o(\Psi^o)^{M-1}\}\, , \non\\
\een
where we used the identity
\ben
\langle \delta\wt\Psi^c|c_0^-Q_B\mathcal G|\Psi^c \rangle
&=&\langle \wt\Psi^c|c_0^-Q_B\mathcal G|\delta\wt\Psi^c \rangle\, , \non\\[.2cm]
\langle \delta\wt\Psi^o|Q_B\mathcal G|\wt\Psi^o \rangle&=&\langle 
\wt\Psi^o|Q_B\mathcal G|\delta\wt\Psi^o \rangle\, .
\een
We shall analyze separately the effect of the transformations generated by 
$\wt \Lambda^c$, $\wt\Lambda^o$, 
$\Lambda^c$ and $\Lambda^o$.
First we consider the gauge transformation parametrized by $\wt\Lambda^c$ in \refb{e1.22}. 
Using $Q_B^2=0$ we see that the variation of the action is given by
\ben
\wt\delta_c S_{1PI} &=&   \{Q_B\wt{\Lambda}^c\}_D \, .
\een
But this vanishes by the identity \refb{emainex}. 
Similarly, the variation of the action parametrized by $\wt\Lambda^o$ in \refb{e1.22}
vanishes identically:
\ben
\wt\delta_o S_{1PI} &=&  0\, . 
\een

For the gauge transformation parametrized by $\Lambda^c$, we have, from \refb{e1.22},
\ben
&&\hskip -.25in \delta_c S_{1PI} % \non\\
=   -\sum_{N\ge 0}\sum_{M\ge 0}\f{1}{N!M!}\langle \wt\Psi^c|c_0^-Q_B\mathcal G| [\Lambda^c(\Psi^c)^N;(\Psi^o)^M]^c \rangle \non\\ &&
 \hskip -.25in +\ \sum_{N\ge 0}\sum_{M\ge 0}\f{1}{N!M!}\langle  [\Lambda^c(\Psi^c)^N;(\Psi^o)^M]^c|c_0^-Q_B|\Psi^c \rangle 
%\non\\[.2cm] && 
+\ \sum_{N\ge 0}\sum_{M\ge 0}\f{1}{N!M!}\langle \wt\Psi^c|c_0^-Q_B\mathcal{G}|[\Lambda^c(\Psi^c)^N;(\Psi^o)^M]^c \rangle \non\\ &&
 \hskip -.25in +\sum_{N\ge 0}\sum_{M\ge 0}\f{1}{N!M!}\langle \wt\Psi^o|Q_B\mathcal G|[\Lambda^c(\Psi^c)^N;(\Psi^o)^M]^o\rangle\ %\non\\ &&
-\ \sum_{N\ge 0}\sum_{M\ge 0}\f{1}{N!M!}\langle [\Lambda^c(\Psi^c)^N;(\Psi^o)^M]^o|Q_B|\Psi^o \rangle\non\\[.2cm]
&& \hskip -.25in -\ \sum_{N\ge 0}\sum_{M\ge 0}\f{1}{N!M!} \langle \wt\Psi^o|Q_B\mathcal G|[\Lambda^c(\Psi^c)^N;(\Psi^o)^M]^o\rangle %\non\\ &&
+g_s^2\sum_{N\ge 0}\sum_{M\ge 0}\f{1}{N!M!} \{[\Lambda^c(\Psi^c)^N;(\Psi^o)^M]^c\}_D  \non\\[.2cm]
&& \hskip -.25in+\sum_{N= 1}^\infty\sum_{M= 0}^\infty\f{1}{(N-1)!M!}\{(\Psi^c)^{N-1}Q_B\Lambda^c;(\Psi^o)^M\} \non\\
&& \hskip -.25in+\sum_{N= 1}^\infty\sum_{M= 0}^\infty\sum_{P\ge0}\sum_{Q\ge0}\f{g_s^2}{(N-1)!M!P!Q!}\{(\Psi^c)^{N-1}\mathcal{G}[\Lambda^c(\Psi^c)^P;(\Psi^o)^Q]^c;(\Psi^o)^M\} \non\\
&& \hskip -.25in -\sum_{N= 0}^\infty\sum_{M= 1}^\infty\sum_{P\ge0}\sum_{Q\ge0}\f{g_s}{N!(M-1)!P!Q!}\{(\Psi^c)^N;\mathcal{G}[\Lambda^c(\Psi^c)^P;(\Psi^o)^Q]^o(\Psi^o)^{M-1}\} \, .
\een
The first and 3rd terms cancel each other while 4th and 6th terms cancel each other. 
After using \refb{edefsq} we are left with
\ben\label{edelcS}
&& \hskip -.25in 
\delta_c S_{1PI} =    \ \sum_{N\ge 0}\sum_{M\ge 0}\f{1}{N!M!}\{ (Q_B\Psi^c)\Lambda^c(\Psi^c)^N;(\Psi^o)^M\} 
\non\\ &&  \hskip -.25in
-\ \sum_{N\ge 0}\sum_{M\ge 0}\f{1}{N!M!}\{\Lambda^c(\Psi^c)^N;(Q_B\Psi^o)(\Psi^o)^M\}
% \non\\[.2cm] && 
+g_s^2\sum_{N\ge 0}\sum_{M\ge 0} \f{1}{N!M!}\{[\Lambda^c(\Psi^c)^N;(\Psi^o)^M]^c\}_D  \non\\ && 
 \hskip -.25in +\sum_{N= 1}^\infty\sum_{M= 0}^\infty\f{1}{(N-1)!M!}\{(\Psi^c)^{N-1}Q_B\Lambda^c;(\Psi^o)^M\} \non\\
&& \hskip -.25in+\sum_{N= 1}^\infty\sum_{M= 0}^\infty\sum_{P\ge0}\sum_{Q\ge0}\f{g_s^2}{(N-1)!M!P!Q!}\{(\Psi^c)^{N-1}\mathcal{G}[\Lambda^c(\Psi^c)^P;(\Psi^o)^Q]^c;(\Psi^o)^M\} \non\\
&& \hskip -.25in -\sum_{N= 0}^\infty\sum_{M= 1}^\infty\sum_{P\ge0}\sum_{Q\ge0}\f{g_s}{N!(M-1)!P!Q!}\{(\Psi^c)^N;\mathcal{G}[\Lambda^c(\Psi^c)^P;(\Psi^o)^Q]^o(\Psi^o)^{M-1}\}\, . 
\een
Now, we specialize the identity \refb{emainsup} to the following case
\ben
A_i^c &=& \Psi^c \qquad \mbox{for} \ \ i= 1,\cdots, N-1\, , \non\\
A_N^c &=&\zeta \Lambda^c\, ,\non\\
 A_j^o &=& \Psi^o\qquad \mbox{for} \ \ j= 1,\cdots, M\, ,
\een
where $\zeta$ is a grassmann odd $c$-number. This gives
\ben
&&(N-1) \bigl\{Q_B\Psi^c(\Psi^c)^{N-2}\zeta\Lambda^c;(\Psi^o)^M\bigl\}+ \bigl\{(\Psi^c)^{N-1}Q_B\zeta\Lambda^c;(\Psi^o)^M\bigl\} \non\\[.3cm]
&&+\ M \bigl\{(\Psi^c)^{N-1}\zeta\Lambda^c;(Q_B\Psi^o)(\Psi^o)^{M-1}\bigl\}\non\\
&=&-\sum_{k=0}^{N-1}\sum_{\ell=0}^M {N-1\choose k}\, {M\choose \ell}\, \Bigl(g_s^2 \bigl\{(\Psi^c)^k\mathcal G[(\Psi^c)^{N-1-k}\zeta\Lambda^c;(\Psi^o)^{M-\ell}]^c;(\Psi^o)^{\ell}\bigl\}\non\\
&&\hspace*{1.2in}+\  g_s\bigl\{(\Psi^c)^k;\mathcal G[(\Psi^c)^{N-1-k}\zeta\Lambda^c;(\Psi^o)^{M-\ell}]^o(\Psi^o)^{\ell}\bigl\}\Bigl)\non\\[.2cm]
&&-g_s^2 \{[(\Psi^c)^{N-1}\zeta\Lambda^c;(\Psi^o)^M]^c\}_D\, .
\een
In writing the above equation, we have used the identities \eqref{idensup1},
\eqref{idensup2}.
Now, bringing the grassmann odd parameter $\zeta$ to the extreme left, multiplying the expression by
$\{(N-1)! M!\}^{-1}$, and summing over $M$ and $N$, we obtain
\ben
0&=&\sum_{M\ge 0}\sum_{N\ge 2}\f{1}{(N-2)!M!} \bigl\{Q_B\Psi^c(\Psi^c)^{N-2}\Lambda^c;(\Psi^o)^M\bigl\}\non\\ &&
+ \sum_{M\ge 0}\sum_{N\ge 1}\f{1}{(N-1)!M!} \bigl\{(\Psi^c)^{N-1}Q_B\Lambda^c;(\Psi^o)^M\bigl\} \non\\[.3cm]
&&-\ \sum_{M\ge 0}\sum_{N\ge 1}\f{1}{(N-1)!(M-1)!}  \bigl\{(\Psi^c)^{N-1}\Lambda^c;(Q_B\Psi^o)(\Psi^o)^{M-1}\bigl\}\non\\
&&+\sum_{M\ge 0}\sum_{N\ge 1} \sum_{k=0}^{N-1}\sum_{\ell=0}^M\f{1}{(N-1-k)!(M-\ell)!\ell!k!}\Bigl(g_s^2 \bigl\{(\Psi^c)^k\mathcal G[(\Psi^c)^{N-1-k}\Lambda^c;(\Psi^o)^{M-\ell}]^c;(\Psi^o)^{\ell}\bigl\}\non\\
&&\hspace*{1.2in}-\  g_s\bigl\{(\Psi^c)^k;\mathcal G[(\Psi^c)^{N-1-k}\Lambda^c;(\Psi^o)^{M-\ell}]^o(\Psi^o)^{\ell}\bigl\}\Bigl)\non\\[.2cm]
&&+g_s^2 \sum_{M\ge 0}\sum_{N\ge 1}\f{1}{(N-1)!M!} \{[(\Psi^c)^{N-1}\Lambda^c;(\Psi^o)^M]^c\}_D\, .
\een
After redefining the sums appropriately, we see that the right hand side of the above equation is precisely the gauge transformation $\delta_c S_{1PI}$ given in 
\refb{edelcS}. This proves gauge invariance under the transformation 
generated by $\Lambda^c$:
\be
\delta_c S_{1PI}=0\, .
\ee

Finally, we consider the variation of the action under the gauge transformation parametrized by $\Lambda^o$. The variation of the action is given by
\ben
&& \hskip -.25in \delta_o S_{1PI} =   -\sum_{N\ge 0}\sum_{M\ge 0}\f{1}{N!M!}\langle \wt\Psi^c|c_0^-Q_B\mathcal G| [(\Psi^c)^N;\Lambda^o(\Psi^o)^M]^c \rangle \non\\ && 
\hskip -.25in +\ \sum_{N\ge 0}\sum_{M\ge 0}\f{1}{N!M!}\langle  [(\Psi^c)^N;\Lambda^o(\Psi^o)^M]^c|c_0^-Q_B|\Psi^c \rangle %\non\\[.2cm] && 
+\ \sum_{N\ge 0}\sum_{M\ge 0}\f{1}{N!M!}\langle \wt\Psi^c|c_0^-Q_B\mathcal{G}|[(\Psi^c)^N;\Lambda^o(\Psi^o)^M]^c \rangle\non\\
&& \hskip -.25in +\sum_{N\ge 0}\sum_{M\ge 0}\f{1}{N!M!}\langle \wt\Psi^o|Q_B\mathcal G|[(\Psi^c)^N;\Lambda^o(\Psi^o)^M]^o\rangle\ %\non\\ &&
-\ \sum_{N\ge 0}\sum_{M\ge 0}\f{1}{N!M!}\langle [(\Psi^c)^N;\Lambda^o(\Psi^o)^M]^o|Q_B|\Psi^o \rangle\non\\[.2cm]
&&\hskip -.25in  -\ \sum_{N\ge 0}\sum_{M\ge 0}\f{1}{N!M!} \langle \wt\Psi^o|Q_B\mathcal G|[(\Psi^c)^N;\Lambda^o(\Psi^o)^M]^o\rangle %\non\\ &&
+g_s^2\sum_{N\ge 0}\sum_{M\ge 0}\f{1}{N!M!} \{[(\Psi^c)^N;\Lambda^o(\Psi^o)^M]^c\}_D  \non\\[.2cm]
&&\hskip -.25in 
+\sum_{N= 0}^\infty\sum_{M= 1}^\infty\f{1}{(M-1)!N!}\{(\Psi^c)^{N};Q_B\Lambda^o(\Psi^o)^{M-1}\} \non\\
&&\hskip -.25in +\sum_{N= 1}^\infty\sum_{M= 0}^\infty\sum_{P\ge0}
\sum_{Q\ge0}\f{g_s^2}{(N-1)!M!P!Q!}\{(\Psi^c)^{N-1}\mathcal{G}[(\Psi^c)^P;
\Lambda^o(\Psi^o)^Q]^c;(\Psi^o)^M\} \non\\
&& \hskip -.25in 
-\sum_{N= 0}^\infty\sum_{M= 1}^\infty\sum_{P\ge0}\sum_{Q\ge0}\f{g_s}{N!(M-1)!P!Q!}
\{(\Psi^c)^N;\mathcal{G}[(\Psi^c)^P;\Lambda^o(\Psi^o)^Q]^o(\Psi^o)^{M-1}\} \, . % \non
\een
Again, the first and third terms cancel each other and the fourth and sixth terms cancel each other.
After using \refb{edefsq}, we are left with
\ben
&& \hskip -.25in \delta_o S_{1PI} =  \ \sum_{N\ge 0}\sum_{M\ge 0}\f{1}{N!M!}\langle  
\{(Q_B\Psi^c) (\Psi^c)^N;\Lambda^o(\Psi^o)^M\} \non\\[.2cm]
&&\hskip -.25in 
-\ \sum_{N\ge 0}\sum_{M\ge 0}\f{1}{N!M!}\{(\Psi^c)^N;(Q_B\Psi^o)\Lambda^o(\Psi^o)^M\}
%\non\\[.2cm]&&
+g_s^2\sum_{N\ge 0}\sum_{M\ge 0}\f{1}{N!M!} \{[(\Psi^c)^N;\Lambda^o(\Psi^o)^M]^c\}_D  \non\\[.2cm]
&&\hskip -.25in
+\sum_{N= 0}^\infty\sum_{M= 1}^\infty\f{1}{(M-1)!N!}\{(\Psi^c)^{N};Q_B\Lambda^o(\Psi^o)^{M-1}\} \non\\
&&\hskip -.25in
+g_s^2\sum_{N= 1}^\infty\sum_{M= 0}^\infty\sum_{P\ge0}\sum_{Q\ge0}
\f{1}{(N-1)!M!P!Q!}\{(\Psi^c)^{N-1}\mathcal{G}[(\Psi^c)^P;\Lambda^o(\Psi^o)^Q]^c;(\Psi^o)^M\} 
\non\\
&& \hskip -.25in
-g_s \sum_{N= 0}^\infty\sum_{M= 1}^\infty\sum_{P\ge0}\sum_{Q\ge0}\f{1}{N!(M-1)!P!Q!}\{(\Psi^c)^N;
\mathcal{G}[(\Psi^c)^P;\Lambda^o(\Psi^o)^Q]^o(\Psi^o)^{M-1}\} \, .\label{op_gaug}
\een
Now, we specialize the identity \refb{emainsup} to the following case
\ben
A_i^o &=& \Psi^o \qquad \mbox{for} \ \ i= 1,\cdots, M-1\, , \non\\
A_N^o &=&\zeta \Lambda^o\, , \non\\
 A_j^c &=& \Psi^c\qquad \mbox{for} \ \ j= 1,\cdots, N \, ,
\een
for some grassmann odd $c$-number $\zeta$. This gives
\ben
&&N \bigl\{Q_B\Psi^c(\Psi^c)^{N-1};\zeta\Lambda^o(\Psi^o)^{M-1}\bigl\}
+ \bigl\{(\Psi^c)^{N};(Q_B\zeta\Lambda^o)(\Psi^o)^{M-1}\bigl\} \non\\[.3cm]
&&+\ (M-1) \bigl\{(\Psi^c)^{N};(Q_B\Psi^o)\zeta\Lambda^o(\Psi^o)^{M-1}
\bigl\}\non\\
&=&-\sum_{k=0}^{N}\sum_{\ell=0}^{M-1}{N\choose k}{M-1\choose \ell}
\Bigl(g_s^2 \bigl\{(\Psi^c)^k\mathcal G[(\Psi^c)^{N-k};\zeta
\Lambda^o(\Psi^o)^{M-\ell-1}]^c;(\Psi^o)^{\ell}\bigl\}\non\\
&&\hspace*{1.2in}+\  g_s\bigl\{(\Psi^c)^k;\mathcal G[(\Psi^c)^{N-1-k};\zeta\Lambda^o(\Psi^o)^{M-\ell-1}]^o(\Psi^o)^{\ell}\bigl\}\Bigl)\non\\[.2cm]
&&-g_s^2 \{[(\Psi^c)^{N};\zeta\Lambda^o(\Psi^o)^{M-1}]^c\}_D\, .
\een
Bringing the grassmann odd parameter $\zeta$ to extreme left and 
summing over $M$ and $N$ after multiplying with $\{N!(M-1)!\}^{-1}$, we get
\ben
0&=&\sum_{N\ge 1}\sum_{M\ge 1}\f{1}{(N-1)!(M-1)!} \bigl\{(Q_B\Psi^c)(\Psi^c)^{N-1};\Lambda^o(\Psi^o)^{M-1}\bigl\}
\non\\ &&
%\hskip -.25in 
+ \sum_{N\ge 0}\sum_{M\ge 1}\f{1}{N!(M-1)!}\bigl\{(\Psi^c)^{N};(Q_B\Lambda^o)(\Psi^o)^{M-1}\bigl\} 
\non\\[.3cm] &&
-\ \sum_{N\ge 0}\sum_{M\ge 2}\f{1}{N!(M-2)!} \bigl\{(\Psi^c)^{N};(Q_B\Psi^o)\Lambda^o(\Psi^o)^{M-2}\bigl\}\non\\
&&+\sum_{N\ge 0}\sum_{M\ge 1}\sum_{k=0}^{N}\sum_{\ell=0}^{M-1}\f{1}{(N-k)!(M-\ell-1)!k!\ell!}\Bigl(g_s^2 \bigl\{(\Psi^c)^k\mathcal G[(\Psi^c)^{N-k};\Lambda^o(\Psi^o)^{M-\ell-1}]^c;(\Psi^o)^{\ell}\bigl\}\non\\
&&\hspace*{1.2in}-\  g_s\bigl\{(\Psi^c)^k;\mathcal G[(\Psi^c)^{N-1-k};\Lambda^o(\Psi^o)^{M-\ell-1}]^o(\Psi^o)^{\ell}\bigl\}\Bigl)\non\\[.2cm]
&&+g_s^2 \sum_{N\ge 0}\sum_{M\ge 1}\f{1}{N!(M-1)!} \{[(\Psi^c)^{N};\Lambda^o(\Psi^o)^{M-1}]^c\}_D\, .
\een
After redefining the sums and comparing with \eqref{op_gaug}, we find 
\ben
\delta_o S_{1PI} =0\, .
\een

This completes the proof of gauge invariance of the 1PI effective action.

\newcommand{\blan}{}

\sectiono{Classical BV master equation for the 1PI action} \label{scbv}

We shall now show that the 1PI effective action satisfies the classical BV master 
equation\cite{bochicchio,thorn,9206084,9705241}. 
For this, we first identify the fields and anti-fields of the theory. This is done by dividing the Hilbert spaces as follows
\ben
\blan \HH^c \ =\ \blan \HH^c_+\ \oplus\ \blan \HH^c_-\, , \qquad\qquad \widetilde \HH^c \ =\ \widetilde \HH^c_+\ \oplus\ \widetilde \HH^c_-\, ,\non\\[.3cm]
\blan \HH^o \ =\ \blan \HH^o_+\ \oplus\ \blan \HH^o_-\, , \qquad\qquad \widetilde \HH^o \ =\ \widetilde \HH^o_+\ \oplus\ \widetilde \HH^o_-\, ,
\een
such that the states in $\blan \HH^c_+$ and $\widetilde \HH^c_+$ have world-sheet ghost number 
$\ge 3$, the states in $\blan \HH^c_-$ and $\widetilde \HH^c_-$ have world-sheet ghost 
number $\le 2$,
 the states in $\blan \HH^o_+$ and $\widetilde \HH^o_+$ have world-sheet ghost number $\ge 2$ 
 and the states in $\blan \HH^c_-$ and $\widetilde \HH^c_-$ have the world-sheet ghost number 
 $\le 1$.
We denote the basis states of $\blan \HH^c_+,\ \blan \HH^c_-,\ \widetilde \HH^c_+,$ and 
$ \widetilde \HH^c_- $ by $|\varphi_+^r\rangle, |\varphi^-_r\rangle , |\tilde \varphi^r_+\rangle $ 
and $|\tilde\varphi^-_r\rangle$ respectively. They are chosen to
satisfy orthonormality and completeness conditions:
\ben
&& \langle\tilde\vp^r_+|c_0^-|\vp_s^-\rangle = \delta^r_s = \langle\vp_s^-|c_0^-|\tilde\vp^r_+\rangle, \quad
\langle\vp^r_+|c_0^-|\tilde\vp_s^-\rangle = \delta^r_s = \langle\tilde\vp_s^-|c_0^-|\vp^r_+\rangle,
\non\\ &&
|\vp_r^-\rangle \langle\tilde\vp^r_+| +  |\vp^r_+\rangle \langle\tilde\vp_r^-|= b_0^- = |\tilde\vp^r_+\rangle \langle\vp_r^-|
+  |\tilde\vp_r^-\rangle \langle\vp^r_+|\, .
\een
Similarly,  we denote the basis states of $\blan \HH^o_+,\ \blan \HH^o_-,\ \widetilde \HH^o_+,$ and 
$ \widetilde \HH^o_- $ by $|\blan\phi_+^r\rangle, |\blan\phi^-_r\rangle , |\tilde\phi^r_+\rangle $ and $|\tilde\phi^-_r\rangle$ respectively. They satisfy orthonormality and completeness conditions:
\ben
&& \langle\tilde\phi^r_+|\phi_s^-\rangle = \delta^r_s = \langle\phi_s^-|\tilde\phi^r_+\rangle, \quad
\langle\phi^r_+|\tilde\phi_s^-\rangle = \delta^r_s = \langle\tilde\phi_s^-|\phi^r_+\rangle,
\non\\ &&
|\phi_r^-\rangle \langle\tilde\phi^r_+| +  |\phi^r_+\rangle \langle\tilde\phi_r^-|= 1
= |\tilde\phi^r_+\rangle \langle\phi_r^-|
+  |\tilde\phi_r^-\rangle \langle\phi^r_+|\, .
\een

The closed string fields are expanded as\footnote{in these equations * denotes anti-fields and
not complex conjugation.}
\ben\label{eclosed}
|\widetilde\Psi^c\rangle \ =\  g_s\, \sum_r\ (-1)^{\tilde \varphi_r^-} ( \widetilde\psi^c)^r 
|\wt\varphi_r^-\rangle- g_s\, \sum_r \  (\psi^c)_r^* \, |\wt\varphi^r_+\rangle\, ,\non\\
|\Psi^c\rangle -\f{1}{2}\mathcal G|\widetilde\Psi^c\rangle\ =\ g_s\, \sum_r\ 
(-1)^{\varphi_r^-}\, 
 (\psi^c)^r |\blan\varphi_r^-\rangle- g_s\, \sum_r \ ( \widetilde\psi^c)_r^*\, |\blan\varphi^r_+\rangle\, ,
\een
where $\vp$ in the exponent for any state $|\vp\rangle$ denotes the grassmann parity of the
vertex operator $\vp$, taking value 0 for even operators and 1 for
odd operators.
We define the target space ghost number of the coefficient fields by $g=2-G$ where $G$ 
denotes the world-sheet ghost number of the corresponding basis states. This means that the 
coefficients $(\psi^c)^r, (\wt\psi^c)^r$ have target space ghost numbers $\ge0$ whereas the coefficients 
$(\psi^c)^*_r$ and $(\wt\psi^c)^*_r$ have target space ghost numbers $\le -1$. 
In the BV 
quantization, the $(\psi^c)^r$ and $(\wt\psi^c)^r$ will be interpreted as fields 
whereas $(\psi^c)^*_r$ and $(\wt\psi^c)^*_r$ will be interpreted as anti-fields. The factors of
$g_s$ in \refb{eclosed} ensure that in the action $(\psi^c)^r$, $(\wt\psi^c)^r$, $(\psi^c)^*_r$ 
and $(\wt\psi^c)^*_r$ have conventionally normalized kinetic terms.

In a similar way, we expand the open string fields as 
\ben\label{eopen}
|\widetilde\Psi^o\rangle \ =\ g_s^{1/2}\, \sum_r\ ( \widetilde\psi^o)^r\,  |\tilde\phi_r^-\rangle 
+ g_s^{1/2}\, \sum_r (-1)^{\tilde\phi^r_++1}\  (\psi^o)_r^*\, |\tilde\phi^r_+\rangle\, ,\non\\
|\Psi^o\rangle -\f{1}{2}\mathcal G|\widetilde\Psi^o\rangle\ =\ g_s^{1/2}\, \sum_r\   (\psi^o)^r \, 
|\blan\phi_r^-\rangle +g_s^{1/2}\, 
 \sum_r (-1)^{\phi^r_++1}\  ( \widetilde\psi^o)_r^*\, |\blan\phi^r_+\rangle\, .
\een
We define the target space ghost number of the coefficient fields by $g=1-G$ where $G$ denotes the 
world-sheet ghost number of the corresponding basis states. This means that the coefficients 
$(\psi^o)^r, (\wt\psi^o)^r$ have target space ghost numbers $\ge0$ whereas the coefficients $(\psi^o)^*_r$ 
and $(\wt\psi^o)^*_r$ have target space ghost numbers $\le -1$. 
In the BV quantization, 
$(\psi^o)^r$ and $(\wt\psi^o)^r$ will be interpreted as fields 
whereas 
$(\psi^o)^*_r$ and $(\wt\psi^o)^*_r$ 
will be interpreted as anti-fields.

The BV anti-bracket between any two functions $F$ and $G$ of the fields is given by 
\ben\label{edefab}
(F,G) = \f{\p_R F}{\p\psi^r} \f{\p_L G}{\p\psi^*_r}- \f{\p_R F}{\p\psi_r^*} \f{\p_L G}{\p\psi^r}\, ,
\een
where $\psi^r$ stand for all the fields $(\psi^c)^r$, $(\wt\psi^c)^r$, $(\psi^o)^r$ and $(\wt\psi^o)^r$
and $\psi^*_r$ stand for all the anti-fields $(\psi^c)^*_r$, $(\wt\psi^c)^*_r$, $(\psi^o)^*_r$ and $(\wt\psi^o)^*_r$.
$\p_L$ and $\p_R$ denotes left and right derivatives respectively.
If for an arbitrary function $F(\Psi^c,\wt\Psi^c,\Psi^o,\widetilde\Psi^o)$, one 
expresses the first order variation as
\ben \label{eanticonpre}
\delta F\ &=&\ \langle F^c_R|c_0^-|\delta\widetilde\Psi^c\rangle\ 
+\ \langle \widetilde F^c_R|c_0^-|\delta\Psi^c\rangle\  + \langle F^o_R|\delta\widetilde\Psi^o\rangle\ 
+\ \langle \widetilde F^o_R|\delta\Psi^o\rangle\ \non\\
&=& \ \langle  \delta\widetilde\Psi^c|c_0^-|F^c_L\rangle\ +\ \langle  \delta\Psi^c|c_0^-|\widetilde F^c_L\rangle+
\ \langle  \delta\widetilde\Psi^o|F^o_L\rangle\ +\ \langle  \delta\Psi^o|\widetilde F^o_L\rangle\label{4.2.6}\, ,
\een
then using \refb{eclosed}, \refb{eopen}, \refb{edefab}, one can express the anti-bracket  as
\ben \label{eanticon}
(F,G) &=& - g_s^2 \Bigl(  \langle F^c_R|c_0^-|\widetilde G^c_L\rangle \ +\ \langle \widetilde F^c_R|c_0^-| G^c_L\rangle \ +\  \langle\widetilde F^c_R|c_0^-\mathcal G|\widetilde G^c_L\rangle\Bigl) \non\\
&&  -\ {g_s}\Bigl(\langle F^o_R|\widetilde G^o_L\rangle \ +\ \langle \widetilde F^o_R| G^o_L\rangle \ +\  \langle\widetilde F^o_R|\mathcal G|\widetilde G^o_L\rangle \Bigl)\, .
\een

Our goal will be to verify that the action \refb{e1.21} satisfies the classical BV master equation:
\ben
(S_{1PI},S_{1PI})= 0\, .
\een
For this we need to compute the quantities $S^{c,o}_R, S^{c,o}_L, 
\widetilde S^{c,o}_R$ and $\widetilde S^{c,o}_L$ using the definition \eqref{4.2.6}. 
The variation of the action \refb{e1.21} under an arbitrary deformation of the fields is given by
\ben
\delta S_{1PI} &=&     
 -\f{1}{g_s^2}\langle \delta\wt\Psi^c|c_0^-Q_B\mathcal G|\wt\Psi^c \rangle 
 +\ \f{1}{g_s^2}\langle \delta\wt\Psi^c|c_0^-Q_B|\Psi^c \rangle \ 
 +\ \f{1}{g_s^2}\langle \delta\Psi^c|c_0^-Q_B|\wt\Psi^c \rangle\non\\
&&-\ \f{1}{g_s}\langle\delta \wt\Psi^o|Q_B\mathcal G|\wt\Psi^o \rangle\ 
+\ \f{1}{g_s}\langle \delta\wt\Psi^o|Q_B|\Psi^o \rangle\ +\ \f{1}{g_s}\langle\delta \Psi^o|Q_B|\wt\Psi^o \rangle\ 
+\ \langle\delta\wt\Psi^c|c_0^-|[\ ]_D\rangle\non\\[.2cm]
&& +\sum_{N= 1}^\infty\sum_{M= 0}^\infty\f{1}{(N-1)!M!}\bigl\langle\delta\Psi^c|c_0^-|[(\Psi^c)^{N-1};(\Psi^o)^M]^c\bigl\rangle \non\\
&& +\sum_{N= 0}^\infty\sum_{M= 1}^\infty\f{1}{N!(M-1)!}\langle\delta\Psi^o|[(\Psi^c)^N;
(\Psi^o)^{M-1}]^o\rangle \, .\label{varleft}
\een
This gives,
\ben
|S^c_L\rangle &=& -\f{1}{g_s^2}Q_B\mathcal G|\wt\Psi^c \rangle \ +\ \f{1}{g_s^2}Q_B|\Psi^c \rangle\ +\ |[\ ]_D\rangle\, ,
\non\\[.2cm]
 |\widetilde S^c_L\rangle &=& \f{1}{g_s^2}Q_B|\wt\Psi^c \rangle \ +\ \sum_{N= 0}^\infty\sum_{M= 0}^\infty\f{1}{N!M!}\bigl|[(\Psi^c)^{N};(\Psi^o)^M]^c\bigl\rangle \, ,\non\\
 |S^o_L\rangle &=& -\f{1}{g_s}Q_B\mathcal G|\wt\Psi^o \rangle \ +\ \f{1}{g_s}Q_B|\Psi^o \rangle\, ,
 \non\\[.2cm]
 |\widetilde S^o_L\rangle &=& \f{1}{g_s}Q_B|\wt\Psi^o \rangle \ +\ \sum_{N= 0}^\infty\sum_{M= 0}^\infty\f{1}{N!M!}\bigl|[(\Psi^c)^{N};(\Psi^o)^M]^o\bigl\rangle \, .
\een
The variation \eqref{varleft} can also be written as
\ben
\delta S_{1PI} 
& = &   
-\f{1}{g_s^2}\langle \wt\Psi^c|c_0^-Q_B\mathcal G|\delta\wt\Psi^c \rangle  +\ \f{1}{g_s^2}\langle \Psi^c|c_0^-Q_B|\delta\wt\Psi^c \rangle \ +\ \f{1}{g_s^2}\langle \wt\Psi^c|c_0^-Q_B|\delta\Psi^c \rangle\non\\
&&-\ \f{1}{g_s}\langle \wt\Psi^o|Q_B\mathcal G|\delta\wt\Psi^o \rangle\ +\ \f{1}{g_s}\langle \Psi^o|Q_B|\delta\wt\Psi^o \rangle\ +\ \f{1}{g_s}\langle \wt\Psi^o|Q_B|\delta\Psi^o \rangle +\ \langle[\ ]_D|c_0^-|\delta\wt\Psi^c\rangle\non\\[.2cm]
&& +\sum_{N= 1}^\infty\sum_{M= 0}^\infty\f{1}{(N-1)!M!}\bigl\langle[(\Psi^c)^{N-1};(\Psi^o)^M]^c|c_0^-|\delta\Psi^c\bigl\rangle \non\\
&& +\sum_{N= 0}^\infty\sum_{M= 1}^\infty\f{1}{N!(M-1)!}\langle[(\Psi^c)^N;(\Psi^o)^{M-1}]^o|\delta\Psi^o\rangle \, .
\hskip .25in 
\label{varright}
\een
This gives
\ben
\langle S^c_R|& =&  \f{1}{g_s^2}\langle \wt\Psi^c|Q_B\mathcal G \ -\ \f{1}{g_s^2}\langle \Psi^c|Q_B \ +\ \langle [\ ]_D|\, ,
\non\\
\langle \widetilde S^c_R|& =&  -\f{1}{g_s^2}\langle \wt\Psi^c|Q_B \  +\sum_{N= 0}^\infty\sum_{M= 0}^\infty\f{1}{N!M!}\bigl\langle[(\Psi^c)^{N};(\Psi^o)^M]^c\bigl| 
\, ,\non\\\non\\
\langle S^o_R|& =&  -\f{1}{g_s}\langle \wt\Psi^o|Q_B\mathcal G \ +\ \f{1}{g_s}\langle \Psi^o|Q_B\, , \non\\
\langle \widetilde S^o_R|& =&  \f{1}{g_s}\langle \wt\Psi^o|Q_B \  +\sum_{N= 0}^\infty\sum_{M= 0}^\infty\f{1}{N!M!}\bigl\langle[(\Psi^c)^{N};(\Psi^o)^M]^o\bigl| \, ,
\een
in the convention that the operators $Q_B$ and $\GG$ act on the right even though they are
to the right of a bra state. 
Using \refb{eanticon}, \refb{edefsq}, \refb{emainex} and  $Q_B^2=0$, we now get
\ben \label{e315a}
- (S_{1PI},S_{1PI}) 
&=&2\sum_{N=0}^\infty \sum_{M=0}^\infty \f{1}{N!M!} \{ Q_B\Psi^c(\Psi^c)^N;(\Psi^o)^M\}\non\\
&&+2\sum_{N=0}^\infty \sum_{M=0}^\infty \f{1}{N!M!} \{(\Psi^c)^N;Q_B\Psi^o(\Psi^o)^M\}\non\\
&&+g_s^2\sum_{N=0}^\infty \sum_{M=0}^\infty\sum_{P=0}^\infty \sum_{Q=0}^\infty \f{1}{N!M!P!Q!}\{ \mathcal G[(\Psi^c)^N;(\Psi^o)^M]^c(\Psi^c)^P;(\Psi^o)^Q\}\non\\
&&+g_s\sum_{N=0}^\infty \sum_{M=0}^\infty\sum_{P=0}^\infty \sum_{Q=0}^\infty \f{1}{N!M!P!Q!}\{ (\Psi^c)^P;\mathcal G[(\Psi^c)^N;(\Psi^o)^M]^o(\Psi^o)^Q\}\non\\
&&+\ 2\, g_s^2\sum_{N=0}^\infty \sum_{M=0}^\infty \f{1}{N!M!} \{ [(\Psi^c)^N;(\Psi^o)^M]^c\}_D\, .
\een
If we specialize the main identity \refb{emainsup} to the case
\ben
A_i^c= \Psi^c\, , \qquad\ i= 1,\cdots, N\, , \non\\
A_i^o= \Psi^o\, , \qquad\ i= 1,\cdots, M\, ,
\een
we obtain
\ben
&&N \bigl\{Q_B\Psi^c(\Psi^c)^{N-1};(\Psi^o)^{M}\bigl\}+ M\bigl\{(\Psi^c)^{N};Q_B\Psi^o(\Psi^o)^{M-1}\bigl\}
 \non\\[.3cm]
&=&-\f{1}{2}\sum_{k=0}^{N}\sum_{\ell=0}^{M}{N\choose k}{M\choose \ell}
\Bigl(g_s^2 \bigl\{(\Psi^c)^k\mathcal G[(\Psi^c)^{N-k};(\Psi^o)^{M-\ell}]^c;(\Psi^o)^{\ell}\bigl\}\non\\
&&\hspace*{1.2in}+\  g_s\bigl\{(\Psi^c)^k;\mathcal G[(\Psi^c)^{N-k};(\Psi^o)^{M-\ell}]^o(\Psi^o)^{\ell}\bigl\}\Bigl)
\non\\[.2cm]
&&-g_s^2 \{[(\Psi^c)^{N};(\Psi^o)^{M}]^c\}_D\,.
\een
Multiplying by $\f{1}{N!M!}$ and summing over $N$ and $M$ from 0 to $\infty$, we get
\ben
0&=&2\sum_{N=0}^\infty \sum_{M=0}^\infty \f{1}{N!M!} \bigl\{Q_B\Psi^c(\Psi^c)^{N};(\Psi^o)^{M}\bigl\}
+ 2\sum_{N=0}^\infty \sum_{M=0}^\infty \f{1}{N!M!}\bigl\{(\Psi^c)^{N};Q_B\Psi^o(\Psi^o)^{M}\bigl\} \non\\[.3cm]
&&+\sum_{N=0}^\infty \sum_{M=0}^\infty \sum_{k=0}^{N}\sum_{\ell=0}^{M}\f{1}{(N-k)!(M-\ell)!k!\ell!}
\Bigl(g_s^2 \bigl\{(\Psi^c)^k\mathcal G[(\Psi^c)^{N-k};(\Psi^o)^{M-\ell}]^c;(\Psi^o)^{\ell}\bigl\}\non\\
&&\hspace*{1.2in}+\  g_s\bigl\{(\Psi^c)^k;\mathcal G[(\Psi^c)^{N-k};(\Psi^o)^{M-\ell}]^o(\Psi^o)^{\ell}\bigl\}\Bigl)
\non\\[.2cm]
&&+2\ g_s^2\sum_{N=0}^\infty \sum_{M=0}^\infty \f{1}{N!M!} \{[(\Psi^c)^{N};(\Psi^o)^{M}]^c\}_D\, .
\een
By redefining the sums in the second line and comparing with the expression of
anti-bracket $(S_{1PI},S_{1PI})$, we see that
\be
(S_{1PI},S_{1PI})=0\,.
\ee
Therefore the 1PI effective action satisfies the classical BV master equation.

The BV formalism also gives a way to derive the gauge transformation laws \refb{e1.22}.  The 
classical BV master action -- and therefore also the 1PI effective action -- 
is known to be invariant under gauge transformations that transform
any function 
$F$ of the string fields $|\Psi^c\rangle$, $|\wt\Psi^c\rangle$, $|\Psi^o\rangle$ and $|\wt\Psi^o\rangle$ 
as\cite{9309027},
\be
\delta F = (F, (S, \Lambda))\, ,
\ee
where $\Lambda$ is any even function of the fields.
Choosing
\be
\Lambda = g_s^{-2} \,\left(  \langle \wt \Psi^c | c_0^-|\Lambda^c\rangle + 
\langle \Psi^c | c_0^- |\wt\Lambda^c\rangle - 
\langle \wt\Psi^c | c_0^- \GG |\wt\Lambda^c\rangle\right) %\non\\ &&
+g_s^{-1} \left( \langle \wt \Psi^o | \Lambda^o\rangle +  
\langle \Psi^o|\wt\Lambda^o\rangle - \langle \wt \Psi^o | \GG|\wt\Lambda^o\rangle \right)\, ,
\ee
we reproduce the gauge transformation laws given in \refb{e1.22}.

\sectiono{Quantum BV master action} \label{sbv}

We can also write down the
quantum BV master action for the combined open closed string field theory following the procedure
reviewed in \cite{1703.06410}.
It is given by
\ben\label{ebvmaster}
S_{BV} &=&  -\f{1}{2g_s^2}\langle \tilde\Psi^c|c_0^-Q_B\mathcal G|\tilde\Psi^c \rangle \ 
+\ \f{1}{g_s^2}\langle \tilde\Psi^c|c_0^-Q_B|\Psi^c \rangle \ -\ \f{1}{2g_s}
\langle \tilde\Psi^o|Q_B\mathcal G|\tilde\Psi^o \rangle\ +\ \f{1}{g_s}\langle \tilde\Psi^o|Q_B|\Psi^o 
\rangle\non\\[.2cm]
&&+ \{\hspace*{-.05in}
\{\tilde\Psi^c\}\hspace*{-.05in}\}_D +\sum_{N= 0}^\infty\sum_{M= 0}^\infty\f{1}{N!M!}\{\hspace*{-.05in}
\{(\Psi^c)^N;(\Psi^o)^M\}\hspace*{-.05in}\} \,,
\een
where $\{\hspace*{-.05in}\{\cdots\}\hspace*{-.05in}\}$, to be defined
shortly,  denotes the contribution to the 
off shell amplitude due to the elementary interaction vertices of superstring field theory. 
$\{\hspace*{-.05in} \{ \wt A^c\}\hspace*{-.05in}\}_D$ is  defined
exactly in the same way as $\{\wt A^c\}_D$.
$\{\hspace*{-.05in} \{ \cdots \}\hspace*{-.05in} \}$
is defined in a way similar to $\{\cdots\}$ given in
\refb{edefcurly}, except that the region of integration $\RR_{g,b,N,M}$ is replaced 
by a smaller region 
$\overline{\mathcal R}_{g,b,N,M}$ defined as follows. 
Recall that we determine $\RR_{g,b,N,M}$ by 
demanding that $\RR_{g,b,N,M}$'s, together with all section segments 
generated from $\RR_{g',b',N',M'}$'s 
by repeated application of hole creation and sewing punctures on different
Riemann surfaces, generate complete generalized sections of $\wt\PP_{g,b,N,M}$'s whose bases
cover the full moduli spaces $\MM_{g,b,N,M}$.
For $\overline{\RR}_{g,b,N,M}$ we make a similar demand, except that we now also
allow sewing two punctures on the same Riemann surface via the sewing relations
\refb{esew1} or \refb{esew2}. 
A systematic procedure for 
constructing the $\overline{\RR}_{g,b,N,M}$'s can be developed along the same lines as for the
$\RR_{g,b,N,M}$'s, as described
in \S\ref{s2}. We begin with the dimension zero $\overline{\RR}_{g,b,N,M}$'s, which can
be taken to be identical to the dimension zero $\RR_{g,b,N,M}$'s, and then begin building higher
dimensional section segments from the lower dimensional ones
by sewing and hole creation operations described in \refb{esew1},
\refb{esew2} and \refb{esew3}. The only difference from the corresponding procedure for the construction of
$\RR_{g,b,N,M}$'s is that we also allow sewing of punctures on the same Riemann surface.
After constructing all the section segments for a given $g,b,N,M$ this way, we `fill the gap' by
$\overline{\RR}_{g,b,N,M}$ so as to generate a full generalized section of $\wt\PP_{g,b,N,M}$.

The quantum BV master action \refb{ebvmaster}, constructed this way, satisfies the quantum BV
master equation:
\be \label{eqbv}
{1\over 2} \, (S_{BV}, S_{BV}) + \Delta S_{BV}=0\, ,
\ee
where the anti-bracket $(,)$ has been defined in \refb{edefab}, and
\be\label{edefDelta}
\Delta S_{BV} = {\p_R\over \p \psi^r}{\p_L S_{BV}\over \p \psi_r^*}\, .
\ee
Here $\psi^r$ stand for all the fields $(\psi^c)^r$, $(\wt\psi^c)^r$, $(\psi^o)^r$ and $(\wt\psi^o)^r$
and $\psi^*_r$ stand for all the anti-fields $(\psi^c)^*_r$, $(\wt\psi^c)^*_r$, $(\psi^o)^*_r$ and $(\wt\psi^o)^*_r$.
$-(S_{BV}, S_{BV})$ is given by the right hand side of \refb{e315a} with $\{~\}$ replaced by
$\{\hspace*{-.05in} \{ ~\}\hspace*{-.05in} \}$. 
Using \refb{edefDelta} one finds that
$\Delta S$ is given by:
\ben
\Delta S&=&-\f{1}{2} \, g_s^2\, \sum_{N=0}^\infty \sum_{M=0}^\infty \f{1}{N!M!}\{\hspace*{-.05in}\{ (\Psi^c)^N\varphi_s\varphi_r;
(\Psi^o)^M      \}\hspace*{-.05in}\} \langle \tilde\varphi^s|c_0^-\mathcal G|\tilde\varphi^r\rangle\non\\
&&-\f{1}{2} \, g_s\, (-1)^{\phi_s}\, \sum_{N=0}^\infty \sum_{M=0}^\infty \f{1}{N!M!}\{\hspace*{-.05in}\{ (\Psi^c)^N;\phi_s\phi_r
(\Psi^o)^M \}\hspace*{-.05in}\} \langle \tilde\phi^s|\mathcal G|\tilde\phi^r\rangle\, ,
\een
where $|\vp_r\rangle$, $\tilde \vp_r\rangle$, $|\phi_r\rangle$ and $|\tilde\phi_r\rangle$ are the basis states in 
$\HH^c$, $\wt\HH^c$, $\HH^o$ and $\wt\HH^o$, normalized according to \refb{ecomp1}, \refb{ecomp2}, and
$(-1)^{\phi_s}$ denotes grassmann parity of the state $|\phi_s\rangle$.
The BV master equation \refb{eqbv} 
can be proved using a modified version of the main identity for $\{\hspace*{-.05in} \{ ~\}\hspace*{-.05in} \}$:
\ben\label{e5.5}
&&\sum_{i=1}^N \{\hspace*{-.05in}\{A_1^c\cdots, A_{i-1}^c(Q_BA_i^c)A_{i+1}^c\cdots A^c_N|A_{1}^o\cdots,A_{M}^o\}\hspace*{-.05in}\} \non\\
&&+\sum_{j=1}^M(-1)^{j-1} \{\hspace*{-.05in}\{A_1^c\cdots A^c_N|A_1^o\cdots A_{j-1}^o(Q_BA_j^o)A_{j+1}^o\cdots A_{M}^o\}\hspace*{-.05in}\}  \non\\
&=&-\f{1}{2}\sum_{k=0}^N\sum_{\{i_1,\cdots,i_k\}\subset \{1,\cdots,N\}}\sum_{\ell=0}^M\sum_{\{j_1,\cdots,j_\ell\}\subset \{1,\cdots,M\}} \Bigl( g_s^2\{\hspace*{-.05in}\{A_{i_1}^c\cdots, A_{i_k}^c\BB^c|A_{j_1}^o\cdots,A_{j_\ell}^o\}\hspace*{-.05in}\} \non\\
&&\hspace*{2in}+\ g_s\{\hspace*{-.05in}\{A_{i_1}^c\cdots, A_{i_k}^c|\BB^oA_{j_1}^o\cdots,A_{j_\ell}^o\}\hspace*{-.05in}\} \Bigl)\non\\
&&-g_s^2\{\hspace*{-.05in}\{[A^c_1\cdots A^c_N;A^o_1\cdots A^o_M]_D\}\hspace*{-.05in}\}-\f{1}{2}\, g_s^2\,
\{\hspace*{-.05in}\{ A_1^c\cdots A^c_N\varphi_s\varphi_r;A_{1}^o\cdots,A_{M}^o      \}\hspace*{-.05in}\} \langle \tilde\varphi^s|c_0^-\mathcal G|\tilde\varphi^r\rangle \non\\
&&-\f{1}{2}\, g_s \, (-1)^{\phi_s}\, \{\hspace*{-.05in}\{ A_1^c\cdots A^c_N; \phi_s\phi_r  
A_{1}^o\cdots,A_{M}^o   \}\hspace*{-.05in}\} \langle \tilde\phi^s|\mathcal G|\tilde\phi^r\rangle\, .
\een
The proof of this follows the same analysis as used in \S\ref{s2} and appendix \ref{sa}
for the proof of \refb{emainsup}. The last two
terms arise due to the fact that $\overline{\RR}_{g,b,N,M}$ has two extra sets of boundaries compared
to $\RR_{g,b,N,M}$, where two bulk punctures or two boundary punctures on a lower 
dimensional $\overline{\RR}_{g',b',N',M'}$ are
sewed via \refb{esew1} or \refb{esew2} with $s=0$.

Given a set of $\overline{\RR}_{g,b,N,M}$'s satisfying the necessary conditions, we can 
construct a set of $\RR_{g,b,N,M}$'s  by sewing
the $\overline{\RR}_{g',b',N',M'}$'s with each other / itself via the sewing operations
\refb{esew1} and / or \refb{esew2}, subject to the constraint that if we omit one 
such operation, the Riemann surface should not become disconnected. This is
precisely the way we build the 1PI amplitudes from elementary vertices using
Feynman diagram, with the sewing playing the role of joining vertices by
propagators. 
$\RR_{g,b,N,M}$'s constructed this way automatically
satisfy the required conditions described in
\S\ref{s2}.

\sectiono{Unoriented open-closed string field theory} \label{sun}

Our construction of the 1PI effective action or BV master action holds for any
superconformal field theory that we use to
compute the correlation functions of vertex operators that enter the definition of
$\{A^c_1\cdots A^c_N; A^o_1\cdots A^o_M\}$ and $\{\wt A^c\}_D$.
Therefore the same construction is valid for any compactification of type IIA or type IIB
superstring field theory involving NSNS background. 
The construction should also generalize to orientifolds
where we have unoriented strings, but there will be a few differences. 
The main difference will be that the definitions
of $\HH^c$, $\wt\HH^c$, $\HH^o$ and $\wt \HH^o$ will automatically 
include projection by the appropriate 
orientifold operation. Therefore the sewing and hole creation operation will also have
this projection operator. For consistency, now we must also include
non-orientable Riemann surfaces in the construction of superstring field theory interaction vertex,
-- if we start with an  oriented Riemann surface and sew two of its punctures with
the orientifold projection inserted, we shall generate a non-orientable
surface.  Together, the oriented and non-orientable surfaces that will be relevant
for us are known as Klein surfaces. Review of the essential results that we shall
need can be found in \cite{zip,9708084,1209.2459} 
and has been summarized in appendix
\ref{sb}.

Due to the inclusion of non-orientable surfaces, we encounter a few differences from the 
corresponding results in oriented open-closed string field theory. 
\begin{enumerate}
\item Off-shell amplitudes now include sum over the moduli spaces of oriented and non-orientable
surfaces. The oriented surfaces are as usual characterized by their genus 
$g$ and the number of
boundaries $b$, while the non-orientable surfaces are characterized by the number of crosscaps $c$ and the  number of
boundaries $b$\cite{zip}, 
with the Euler character given by $b+c-2$. Therefore in the definition \refb{edefcurly} of 
$\{\cdots\}$, the contribution from the non-orientable surfaces must be weighted by a factor of
$(g_s)^{c+b-2}$, whereas the contribution from the oriented surfaces continue to be weighted by
$(g_s)^{2g + b -2}$. We shall often express such factors as $(g_s)^{2g + b+c -2}$, with the understanding
that oriented surfaces will have $c=0$ and non-orientable surfaces will have $g=0$.
\item We shall now show that in the definition of $\{\cdots\}$ given in
\refb{edefcurly} we must also include an extra factor of
\be \label{eofactor}
2^{-g - (c+b)/2 + M/4}\, ,
\ee
if we normalize 
$\Omega_{g,b,c,N,M}$ in the same way as in the case of oriented string theories, {\it e.g.} for
$c=0$, $\Omega$ is defined with the same normalization as for oriented string theory. 
As we shall explain below, this extra factor \refb{eofactor} 
is needed to ensure that amplitudes
factorize correctly near degeneration.\footnote{The presence of the
$M$ dependent factor may be understood as follows. Since we have a 
factor of $2^{-b/2}$,
and since the open string kinetic term involves a disc amplitude, it would
be natural to multiply the open string kinetic term by a factor of $1/\sqrt 2$. However we can remove
this factor by scaling each open string field by $2^{1/4}$. This introduces the factor of
$2^{M/4}$ in the definition of the interaction terms. There is no such factor for closed
strings since the kinetic term is a genus 0 amplitude, and for this there is no
additional factor of 1/2.}

First let us consider the effect of sewing two bulk punctures on an oriented
surface. Now 
the sewing has a projection operator $P=(1+W)/2$ where $W$
is the operation of world-sheet orientation reversal, possibly accompanied by some 
action on the 
space-time. Insertion of 1 corresponds to the usual sewing via \refb{esew1}
and produces 
a handle. The resulting Riemann surface is an oriented Riemann surface with 
two less closed string punctures and one additional genus compared to the
original Riemann surface. $b$,  $M$  and $c(=0)$ remain fixed under this
operation. Under this change  \refb{eofactor} picks up a factor
of 1/2. This correctly
accounts for the factor of 1/2 that appears in the projection operator $(1+W)/2$.

On the other hand insertion of $W$ changes one of the local coordinates in 
\refb{esew1} to its complex conjugate. Therefore the sewing relation takes the form
\be \label{esew4}
z\bar w=e^{-s-i\theta}\, .
\ee
This is known as a cross handle. If the original Riemann surface had genus $g$,
then this operation produces a non-orientable Klein surface with $2g+2$
crosscaps. $M$ and $b$ remain unchanged. It is easy to verify that \refb{eofactor}
changes by a factor of 1/2 under this operation. This again correctly accounts for the
factor of 1/2 in the projection operator.

If the original surface was non-orientable, with $c$ crosscaps, then the effect of
sewing two
of its bulk punctures may  be analyzed in a similar manner. We again have the
projection operator $(1+W)/2$ with 1 corresponding to sewing with a handle and
$W$ corresponding to sewing with a cross handle. Both operations increase the
number of crosscaps by 2, leaving fixed $g(=0)$, $b$ and $M$. Under this 
\refb{eofactor} picks a factors of 1/2, correctly accounting for the 1/2 in the
projection operator.

Next consider the effect of sewing  a pair of boundary punctures of a Klein surface.
Let us for definiteness consider the case where the two punctures lie
on the same boundary.
Again sewing introduces a factor of
$(1+W)/2$. 
For the term proportional to 1, the sewing reduces the
number $M$ of open string punctures by 2 and increases the number $b$ of boundaries by 1. Under such changes,
\refb{eofactor} picks up a factor of 1/2, correctly accounting for the factor of
1/2 in the projection operator. On the other hand if we pick the term proportional to 
$W$,
then this reduces the number $M$ of open string punctures by 2 and increases the number of crosscaps by 1, 
leaving the number of boundaries unchanged.
Under such a change also \refb{eofactor} picks up the desired factor of 1/2. Similar analysis can be done for the sewing 
of a pair of boundary punctures lying on two different boundaries.

It is easy to verify that \refb{eofactor} is also compatible with separating type degenerations. For example when we sew
two oriented surfaces along bulk punctures, the total numbers for 
$g,c,b$ and $M$ all remain unchanged and
therefore there is no change in \refb{eofactor}. On the other hand in this case even 
though sewing introduces
a factor of $(1+W)/2$, both 1 and $W$ produce the same set of
oriented surfaces and therefore there is no factor of 1/2. Similar agreement can be shown for sewing of boundary punctures.
A description of different kinds of sewing that can arise in unoriented 
open closed string field
theory can be found in \cite{9708084} and reviewed in appendix \ref{sb}. 

\item Finally, in defining $\{\wt A^c\}_D$ we must now include not only one point function
of $c_o^-\wt A^c$ on the disc, but also on $RP^2$, together with an insertion of
$\wh \GG$ as before. The latter 
is similar to one point function on the disc, but with the boundary
condition on the disc replaced by a crosscap. 
In keeping with our discussion above we must accompany each crosscap and disc
by a factor of $1/\sqrt 2$ by including it in the definition of
 $\{\wt A^c\}_D$.  
 
\item The choice of local coordinates at the punctures and PCO locations must 
be compatible with the orientifold projection.

\item In the presence of one or more crosscaps, only the total picture number is conserved but the
holomorphic and anti-holomorphic picture numbers are not separately conserved. Therefore the
corresponding $\wt\PP_{0,b,c,N,M}$ will have multiple branches. We can jump between the branches 
by moving the PCOs to the crosscap via vertical segments and converting holomorphic PCOs into
anti-holomorphic PCOs or vice versa using the boundary condition on the crosscap.

\end{enumerate}

With these few changes,
the construction of $\RR_{g,b,c,N,M}$  for 
unoriented open-closed string field theory proceeds in the same way
as in the case of oriented strings, beginning with zero dimensional 
$\RR_{g,b,c,N,M}$'s. Eqs.\refb{edefcurly} and \refb{edefsqd} are generalized to:
\ben \label{edefcurlyunor}
\{ A^c_1\cdots A^c_N; A^o_{1} \cdots A^o_{M}\} 
&\equiv& \sum_{g,b,c,N,M\ge 0\atop (g,b,c,N,M)\ne (0,1,0,1,0),
(0,0,1,1,0)} \, (g_s)^{2g-2+b+c} \, 2^{-g - (b+c)/2 + M/4}\nonumber \\ &&%\hskip .2in
\int_{\RR_{g,b,c,N,M}} \, \Omega^{g,b,c,N,M}_{6g-6+3b+3c+2N+M}
(A^c_1,\cdots, A^c_N;A^o_1,\cdots, A^o_M)\, ,
\een
and,
\be \label{edefsqdunor}
[]_D = {1\over \sqrt 2}\, {\bf P} \, \wh\GG \, \left\{ e^{-\beta_b (L_0+\bar L_0)} 
|B\rangle + e^{-\beta_c (L_0+\bar L_0)} |C\rangle \right\}\, \, ,
\ee
where $\beta_b$ and $\beta_c$ are positive constants and $|B\rangle$ and $|C\rangle$
are the boundary states for the disc and the crosscap.
The form of the action, various identities described in
\S\ref{s1}, the gauge transformation laws and the definition of the anti-bracket
remains the same. We can also construct quantum BV master action by replacing 
$\RR_{g,b,c,N,M}$'s  by $\overline{\RR}_{g,b,c,N,M}$'s  that satisfy slightly different constraints as described in
\S\ref{sbv}.

\bigskip

{\bf Acknowledgement:} 
We would like to thank Theodore Erler, Yuji Okawa and Barton Zwiebach for useful discussions, and Theodore Erler
and Barton Zwiebach for very useful comments on an earlier version of the manuscript.
We thank the Galileo Galilei Institute for Theoretical Physics and INFN for hospitality and partial support during the workshop "String Theory from a world-sheet perspective" where part of this work has been done.
The work of A.S. was
supported in part by the 
J. C. Bose fellowship of 
the Department of Science and Technology, India and the Infosys chair professorship. 
The visit of A.S. to the Galileo Galilei institute 
was supported by a grant from the Simons Foundation 4036 
341344 AL.

\appendix

\sectiono{Signs of the terms in the `main identity'} \label{sa}

In this appendix we shall determine the signs of various terms on the right hand side of the main identity
\refb{emainsupzeta}, following the strategy outlined at the end of \S\ref{s2}.
We shall do the analysis iteratively in the 
dimension of $\RR_{g,b,N,M}$ (or equivalently $\MM_{g,b,N,M}$) associated with the 
interaction vertex, which we
shall simply refer to as the dimension of the interaction vertex. Therefore we shall assume that
\refb{emainsupzeta} holds for vertices carrying dimension $\le K$ and prove that it holds for vertices of
dimension $K+1$.

\begin{figure}
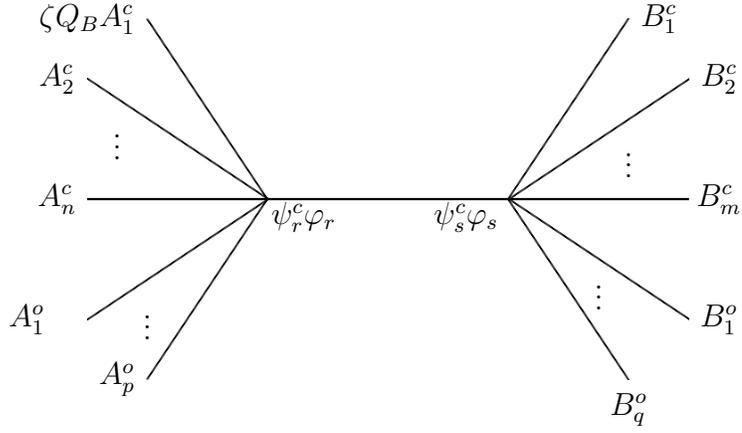


\begin{center}

\figone

\end{center}
\caption{A Feynman diagram where a 1PI vertex
with 
the external states $\zeta Q_B A^c_1,A_2^c\cdots, A^c_n, A^o_1\cdots A^o_p$ and an internal closed string state
and another 1PI vertex with the external
states
$B^c_1,\cdots,  B^c_m, B^o_1,\cdots B^o_q$ and an internal closed string state 
are joined by a closed string propagator.
\label{f1}}

\end{figure}

In the following we shall analyze the behaviour of the amplitude near the boundary of the
moduli space using the language of string field theory. 
This may give the impression that our argument is circular, i.e.\ we use string field theory to prove relations
among string field theory interaction vertices.
However the factorization property of the string amplitude that we use
is known to hold independently of string field theory,
and tells us that the contribution to an amplitude near a separating type degeneration
has the interpretation of the contribution from a Feynman diagram where two different diagrams are
connected by an internal line. This has been shown in Fig.~\ref{f1}.
Therefore our argument does not use any essential element of string field theory.
We shall also use 
gauge invariance of the amplitude, that tells us that if in an amplitude we act $\zeta Q_B$ in turn on
each external state, the result vanishes.
This property of the amplitude also follows from standard world-sheet analysis and does not rely on the
existence of an underlying string field theory.

Let us consider an amplitude with external closed string states $ A^c_1,\cdots A^c_n,
B^c_1,\cdots  B^c_m$ and external open string states 
$A^o_1,\cdots, A^o_p, B^o_1,\cdots, B^o_q$, with $\zeta Q_B$ acting in turn on each
external state. 
Gauge  invariance requires this amplitude to vanish. However the contribution to this 
amplitude from the
1PI vertex does not vanish -- rather it is given by the right hand side of the 
main identity \refb{emainsupzeta}.\footnote{As in \cite{1703.06410} 
we shall follow the convention
that the 1PI vertex contributes to the amplitude without any sign factor.}
Let us first focus on the particular contribution:
\be \label{eyx1}
-g_s^2\, \{A^c_1 \cdots A^c_n \, \GG \zeta
[B^c_1\cdots B^c_m ; B^o_1\cdots B^o_q]^c; A^o_1\cdots A^o_p\}\, ,
\ee
on the right hand side of \refb{emainsupzeta}. This is expected to
get cancelled against the contribution from the Feynman diagram where a 1PI vertex
with 
the external states $A^c_1,\cdots, A^c_n, A^o_1\cdots A^o_p$ and an internal closed string state
and another 1PI vertex with the external
states
$B^c_1,\cdots,  B^c_m, B^o_1,\cdots B^o_q$ and an internal closed string state 
are joined by a closed string propagator,
and $\zeta Q_B$ acts in turn on each of the $A^{c,o}_i$'s and $B^{c,o}_i$'s.
Fig.\ref{f1} shows one of these Feynman diagrams where $\zeta Q_B$ acts on $A_1^c$.
Our strategy will be to evaluate these Feynman diagrams explicitly and  use this to 
test  the sign in \refb{eyx1}. 

We shall first compute 
the contribution to the amplitude from the Feynman diagram when $\zeta Q_B$ acts on $A_1^c$, as shown
in Fig.~\ref{f1}.
Let us denote by $|\vp_r\rangle$ the basis states in $\HH^c$ and by
$|\tilde\vp^r\rangle$ the conjugate basis in $\wt\HH^c$, satisfying
\be \label{ecomp1}
\langle\tilde\vp^r|c_0^-|\vp_s\rangle = \delta^r_s = \langle\vp_s|c_0^-|\tilde\vp^r\rangle,
\quad |\vp_r\rangle \langle\tilde\vp^r| = b_0^- = |\tilde\vp^r\rangle \langle\vp_r|\, ,
\ee
so that the closed string field can be expanded as $\psi_r^c |\vp_r\rangle$.
Then the product of the two vertex factors in Fig.~\ref{f1} can be expressed as:
\be\label{eyy10}
\{(\zeta Q_B A^c_1)\cdots A^c_n \, \psi_r^c\vp_r; A^o_1\cdots A^o_p\} 
\{\psi_s^c \vp_s \, B^c_1\cdots B^c_m ; B^o_1\cdots B^o_q\}\, .
\ee
We shall eventually get a propagator by contracting $\psi_r^c$ and $\psi_s^c$. Therefore $\psi_r^c$ and
$\psi_s^c$ must have same grassmann parity. Since the closed string field is 
grassmann even, $|\vp_r\rangle$ and
$|\vp_s\rangle$ will also have the same grassmann parities as $\psi_r^c$ and $\psi_s^c$. We shall denote this
grassmann parity by $(-1)^{\vp_r}$.
We now
pull the $\psi^c_r$ from the 
$\psi_r^c |\vp_r\rangle$ factor inside the first vertex and 
$\psi^c_s$ from the 
$\psi_s^c |\vp_s\rangle$ factor inside the second vertex to the left
outside the respective $\{\cdots\}$'s. This does not generate any sign since the
$A^c_i$'s are grassmann even.
$\psi^c_r$ and $\psi^c_s$  are now separated by 
$\{(\zeta Q_B A^c_1)\cdots A^c_n \vp_r; A^o_1\cdots A^o_p\}$.
This
has grassmann parity $(-1)^{\vp_r}$.
Therefore the contraction of $\psi^c_r$ and $\psi^c_s$ 
gives a factor of $(-1)^{\vp_r} \Delta_{rs}$ where $\Delta_{rs}$ is the propagator. 

$\Delta_{rs}$ can be computed as follows.
In the kinetic term, we expand 
$\langle\Psi^c|$ as $\psi^c_r \langle \vp_r|$ and pull $\psi^c_r$ to the left without picking any sign, and
express $|\Psi^c\rangle$ as $\psi^c_s |\vp_s\rangle$ and pull $\psi^c_s$ to the left, 
picking a factor of
$(-1)^{\vp_r}$ since it has to pass through $\langle \vp_r| c_0^-Q_B$ which has the same grassmann parity as
$\psi_s^c$. Similar operation can be done for fields in $\wt\HH^c$.
Furthermore in the Siegel gauge we can replace $Q_B$ by
$(c_0L_0+\bar c_0\bar L_0)$.
Inverting the kinetic operator we now get a $\psi_r$-$\psi_s$ 
propagator $\Delta_{rs} = -g_s^2 (-1)^{\vp_r} \langle \tilde\vp^r |c_0^- b_0^+ \GG (L_0^+)^{-1}|\tilde\vp^s
\rangle$\cite{1703.06410}.\footnote{In computing the propagator, we only use the terms involving $Q_B$ as 
kinetic operator and treat the quadratic and linear terms in $\{\cdots\}$ as interaction terms. This allows us
to make contact with the world-sheet formulation, although in order to correctly compute the renormalized 
masses we have to work with the full quadratic term. 
In \cite{1703.06410} the propagator was written as
$-g_s^2 \langle \tilde\vp^s |c_0^- b_0^+ \GG(L_0^+)^{-1}|\tilde\vp^r\rangle$. This can be shown
to be equal to $-g_s^2 (-1)^{\vp_r} \langle \tilde\vp^r |c_0^- b_0^+\GG (L_0^+)^{-1}|\tilde\vp^s\rangle$.}
After cancelling the 
$(-1)^{\vp_r}$ factor in $\Delta_{rs}$ with  the 
$(-1)^{\vp_r}$ factor coming from the contraction of $\psi_r$ and $\psi_s$
across $\{(\zeta Q_B A^c_1)\cdots A^c_n \vp_r; A^o_1\cdots A^o_p\} $,  we arrive at the following
expression for the amplitude shown in Fig.~\ref{f1}:
\be\label{eyy1}
-g_s^2\, \{(\zeta Q_B A^c_1)\cdots A^c_n \vp_r; A^o_1\cdots A^o_p\} \langle \tilde\vp^r |c_0^- b_0^+ \GG\, (L_0^+)^{-1}|\tilde\vp^s\rangle \{\vp_s B^c_1\cdots B^c_m ; B^o_1\cdots B^o_q\}\, .
\ee

We now use \refb{edefsq} to express the last factor as 
$\langle \vp_s|c_0^-|[B^c_1\cdots B^c_m ; B^o_1\cdots B^o_q]^c\rangle$ and then
perform the sum over $s$ in \refb{eyy1} using the completeness relation to express the product of the last two
factors as $\langle\tilde\vp^r |c_0^- b_0^+ \GG\, (L_0^+)^{-1}|[B^c_1\cdots B^c_m ; B^o_1\cdots B^o_q]^c\rangle$.
On the other hand the first $\{\cdots\}$ factor in \refb{eyy1} can be expressed as 
\be
\langle\vp_r|c_0^- [(\zeta Q_B A^c_1)\cdots A^c_n; A^o_1\cdots A^o_p]^c\rangle 
= \langle [(\zeta Q_B A^c_1)\cdots A^c_n; A^o_1\cdots A^o_p]^c|c_0^- |\vp_r\rangle\, ,
\ee
where we have used the fact that $[(\zeta Q_B A^c_1)\cdots A^c_n; A^o_1\cdots A^o_p]^c$ is grassmann odd.
We can now perform the sum over $r$ in \refb{eyy1} using completeness relation and express \refb{eyy1} as
\ben\label{eyy3}
&& -g_s^2 \left\langle \left[(\zeta Q_B A^c_1)\cdots A^c_n ; A^o_1\cdots A^o_p
\right]^c
\right|c_0^-
b_0^+ \GG\, (L_0^+)^{-1}\left|\left[B^c_1\cdots B^c_m ; B^o_1\cdots B^o_q\right]^c
\right\rangle \nonumber \\
&=& -g_s^2 \left\langle b_0^+ \GG\, (L_0^+)^{-1}\left[B^c_1\cdots B^c_m ; B^o_1\cdots B^o_q\right]^c \right| c_0^-\left|
\left[(\zeta Q_B A^c_1)\cdots A^c_n; A^o_1\cdots A^o_p\right]^c\right\rangle\nonumber \\
&=&  - g_s^2 \{(\zeta Q_B A^c_1)\cdots A^c_n \, b_0^+ \GG (L_0^+)^{-1} 
[B^c_1\cdots B^c_m ; B^o_1\cdots B^o_q]^c; A^o_1\cdots A^o_p\}\, ,
\een
where in the first step we have used the fact that both factors of $[\cdots]^c$ 
are grassmann odd and in the last step we have used \refb{edefsq}.

Similar expressions can be obtained when $\zeta Q_B$ acts in turn on the other $A_i^c$'s
and $A_i^o$'s.
Using the main identity for lower number of external states, which we are allowed to use
in a recursive proof, 
the effect of acting with $\zeta Q_B$ in turn on the $B^{c,o}_i$'s can be 
represented by $\zeta Q_B$ acting on $[B^c_1\cdots B^c_m ; B^o_1\cdots B^o_q]^c$.
There are additional terms corresponding to the right hand side of the main identity
with $B^{c,o}_i$'s as external states but these would cancel against other Feynman diagrams
with two propagators, where the right vertex of Fig.~\ref{f1} is replaced by sub-diagram containing an
additional propagator. This brings the sum of the Feynman diagrams of the form shown in
Fig.~\ref{f1} to a form where $\zeta Q_B$ acts in turn
on all the $A^{c,o}_i$'s and on $[B^c_1\cdots B^c_m ; B^o_1\cdots B^o_q]^c$ in the right hand side of
\refb{eyy3}. We again use the main identity on this expression and throw away the right hand
side which would cancel against contribution from Feynman diagrams  where the
left vertex in Fig.~\ref{f1} will be replaced by sub-diagrams with an additional propagator. The
left over term is
given by
$-\zeta Q_B$ acting on $b_0^+ \GG (L_0^+)^{-1}$ in \refb{eyy3}, 
producing just a factor of $-\GG\zeta$.
This brings the relevant contribution from the Feynman diagram with one
closed string propagator
to the form:
\be g_s^2 \{A^c_1 \cdots A^c_n \, \GG \zeta
[B^c_1\cdots B^c_m ; B^o_1\cdots B^o_q]^c; A^o_1\cdots A^o_p\}\, .
\ee
This cancels \refb{eyx1}, confirming that the sign in \refb{eyx1} is correct.

Next we consider the contribution:
\be \label{eyx2}
- g_s \{A^c_1 \cdots A^c_n; \GG\zeta
[B^c_1\cdots B^c_m ; B^o_1\cdots B^o_q]^o A^o_1\cdots A^o_p\}\, ,
\ee
arising from the right hand side of \refb{emainsupzeta}. In the Ward identity for the amplitude,
this is expected to
get cancelled against the contribution from the sum of Feynman diagrams of the same type as the
ones in Fig.~\ref{f1}, except that the internal line is now an open string instead of a closed string.
To evaluate the contribution from the Feynman diagram we first introduce conjugate pair of
basis states
 $|\phi_r\rangle\in\HH^o, |\tilde\phi^r\rangle\in\wt\HH^o$, satisfying
\be\label{ecomp2}
\langle\tilde\phi^r|\phi_s\rangle = \delta^r_s = \langle\phi_s|\tilde\phi^r\rangle,
\quad |\phi_r\rangle \langle\tilde\phi^r| = 1= |\tilde\phi^r\rangle \langle\phi_r|\, .
\ee
Let us first consider the case where $\zeta Q_B$ acts on $A_1^c$.
We proceed as in the earlier case, by first writing down the product of the two vertex factors:
\be\label{ezz10}
\{(\zeta Q_B A^c_1)\cdots A^c_n ; \psi_r^o\phi_r\,  A^o_1\cdots A^o_p\} 
\{ B^c_1\cdots B^c_m ; \psi_s^o \phi_s \, B^o_1\cdots B^o_q\}\, .
\ee
Following the same approach as in the case of \refb{eyy10} we can prove that $\phi_r$,
$\phi_s$ have the same grassmann parity $(-1)^{\phi_r}$, the open string fields $\psi_r^o$ and 
$\psi_s^o$ multiplying them have grassmann parity $(-1)^{\phi_r+1}$, and 
$\{(\zeta Q_B A^c_1)\cdots A^c_n;  \phi_r A^o_1\cdots A^o_p\}$ and 
$\{B^c_1\cdots B^c_m ; \phi_s B^o_1\cdots B^o_q\}$ have grassmann parity $(-1)^{\phi_r+1}$.
We can pull the $\psi^o_r$ and $\psi^o_s$ factors from the two 
amplitudes in \refb{ezz10} to the left outside $\{\cdots\}$ without 
picking any sign. The contraction of $\psi^o_r$ with $\psi^o_s$
produces a factor of $(-1)^{\phi_r+1}$ besides the propagator, since they are separated by
$\{(\zeta Q_B A^c_1)\cdots A^c_n;  \phi_r A^o_1\cdots A^o_p\}$.
The
propagator 
is obtained as follows.
If we express $\langle\Psi^o|$ and 
$|\Psi^o\rangle$ (and similarly $\langle\wt\Psi^o|$) and 
$|\wt\Psi^o\rangle$) in the kinetic term
as $\psi_r^o \langle\phi_r|$ and $\psi_s^o|\phi_s\rangle$ and
then pull $\psi_r^o$ and $\psi_s^o$ to the left, we get a factor of $(-1)^{\phi_r+1}$ from having to pass
$\psi_s^o$ through $\langle\phi_r|Q_B$. 
After replacing $Q_B$ by $c_0L_0$ in the Siegel gauge and inverting the kinetic operator, we get a
$\psi_r^o$-$\psi_s^o$ propagator of the form $-g_s (-1)^{\phi_r+1} \langle \tilde\phi^r |b_0
\GG (L_0)^{-1}|\tilde\phi^s\rangle$. 
Cancelling the $(-1)^{\phi_r+1}$ factor from the propagator with the 
$(-1)^{\phi_r+1}$ factor coming from the contraction, we arrive at the expression:
\be\label{ezz1}
-g_s \, \{(\zeta Q_B A^c_1)\cdots A^c_n;  \phi_r A^o_1\cdots A^o_p\} \langle \tilde\phi^r |b_0
\GG (L_0)^{-1}|\tilde\phi^s\rangle \{B^c_1\cdots B^c_m ; \phi_s B^o_1\cdots B^o_q\}\, .
\ee

The rest of the analysis proceeds almost in the same way as for \refb{eyy1}, with the $[\cdots]^c$'s replaced
by $[\cdots]^o$, $c_0^-$'s removed, $b_0^+$ replaced by $b_0$ and 
$\vp_r$, $\tilde\vp_r$ replaced by $\phi_r$, $\tilde\phi_r$. 
Using the fact that 
$[(\zeta Q_B A^c_1)\cdots A^c_n; A^o_1\cdots A^o_p]^o$ is grassmann even, we can
verify that there are no extra signs compared to those that appeared in the earlier
analysis.
The final result takes the form analogous to \refb{eyy3}
\be \label{ezz3}
 - g_s\{(\zeta Q_B A^c_1)\cdots A^c_n; b_0\GG(L_0)^{-1} 
[B^c_1\cdots B^c_m ; B^o_1\cdots B^o_q]^o A^o_1\cdots A^o_p\} \, .
\ee
We now sum over terms where $\zeta Q_B$ acts in turn on all the external states. Each of these terms
can be analyzed in the same way, leading to expressions similar to \refb{ezz3} with the position of 
$\zeta Q_B$ shifted. Using the main identity for vertices of lower dimension and throwing away terms
that would cancel with Feynman diagrams with two propagators, we can, as in the case of \refb{eyy3}, 
express the sum of these terms as the negative of the term where 
$\zeta Q_B$ acts on $b_0 \GG (L_0)^{-1}$, producing just a factor of $\GG\zeta$.
This brings the boundary contribution to the form:
\be 
g_s\{A^c_1 \cdots A^c_n; \GG\zeta
[B^c_1\cdots B^c_m ; B^o_1\cdots B^o_q]^o A^o_1\cdots A^o_p\}\, .
\ee
This cancels \refb{eyx2}, confirming that the sign in \refb{eyx2} is correct.

\begin{figure}
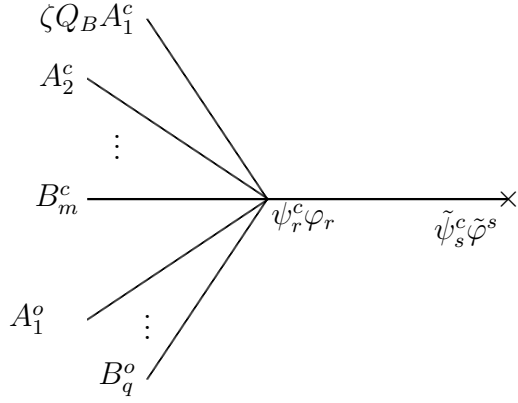


\begin{center}

\figtwo

\end{center}
\caption{A Feynman diagram where a 1PI vertex
with 
the external states $\zeta Q_B A^c_1,A_2^c\cdots, A^c_n, B^c_1,\cdots,  B^c_m,
A^o_1\cdots A^o_p, B^o_1,\cdots B^o_q$ and an internal closed string state in $\HH^c$
and another one point 1PI vertex with an internal closed string state in $\wt\HH^c$
are joined by a closed string propagator.
\label{f2}}

\end{figure}

Finally we turn to the contribution
\be\label{elastcon}
- g_s^2 \{ \zeta[A_c^1\cdots A^c_n B^c_1\cdots B^c_m; A^o_1\cdots A^o_m
B^o_1\cdots B^o_q]^c\}_D\, ,
\ee
arising on the right hand side of \refb{emainsupzeta}. The Feynman diagram that
cancels it is the sum of the diagrams of the form shown in Fig.~\ref{f2}, 
with $\zeta Q_B$ acting on each external state
in turn.
This amplitude can be analyzed in the same way as
\refb{eyy10} and is given by an expression similar to \refb{eyy1}:
\be\label{eww0}
- g_s^2 \{\zeta Q_BA_c^1\cdots A^c_n B^c_1\cdots B^c_m \vp_r; A^o_1\cdots A^o_m
B^o_1\cdots B^o_q\}  \langle \tilde\vp^r |c_0^- b_0^+ \, (L_0^+)^{-1}|\vp_s\rangle
\{\tilde\vp^s\}_D\, ,
\ee
together with similar terms where $\zeta Q_B$ acts in turn on the other external states.
Note that there is no $\GG$ insertion in the propagator since we have to use the $\psi_r$-$\tilde\psi_s$
propagator\cite{1703.06410}. We now carry out manipulations similar to those for \refb{eyy1} to express 
\refb{eww0} in the
form given in \refb{eyy3}:
\be \label{eww3}
- g_s^2 \, \{\zeta Q_BA_c^1\cdots A^c_n B^c_1\cdots B^c_m 
b_0^+ \, (L_0^+)^{-1}[~]_D; A^o_1\cdots A^o_m
B^o_1\cdots B^o_q\} \, .
\ee
The vertex appearing in \refb{eww3} has one less dimension compared to the original vertex, since we replace
a boundary by a closed string puncture where $b_0^+ \, (L_0^+)^{-1}[~]_D$ is 
inserted. Therefore we can use the main identity
\refb{emainsupzeta}. Throwing away the terms on the right hand side
of \refb{emainsupzeta} that would cancel against Feynman 
diagrams with two propagators, and using BRST invariance of $[~]_D$, we can express the sum of 
Feynman diagrams of the type shown in Fig.~\ref{f2} as 
\ben
&& g_s^2 \, \{A_c^1\cdots A^c_n B^c_1\cdots B^c_m 
\zeta[~]_D; A^o_1\cdots A^o_m
B^o_1\cdots B^o_q\} \non\\
&=&
g_s^2 \, \{\zeta [A_c^1\cdots A^c_n B^c_1\cdots B^c_m;A^o_1\cdots A^o_m
B^o_1\cdots B^o_q]^c\}_D
\, ,
\een
where in the last step we used \refb{idensup3}. This cancels \refb{elastcon}, confirming that the sign in
\refb{elastcon} is correct.

\sectiono{Review of non-orientable surfaces} \label{sb}

In this appendix we shall review some properties of non-orientable surfaces
following \cite{zip,9708084,1209.2459}.

A 2 dimensional surface is non-orientable if we cannot assign an orientation 
to the surface uniquely, i.e.\ there exist closed curves such that the tangent bundle,
parallel transported along the curve, comes back with opposite orientation. 
A non-orientable 2 dimensional surface can be described 
using the  `crosscap' --  a disc whose 
diametrically opposite points have been identified. By attaching an arbitrary number of 
crosscaps to a sphere with holes, we can generate an arbitrary non-orientable surface 
in 2 dimensions. Examples of  non-orientable surfaces which appear at tree and one 
loop level in unoriented string theory are real projective plane (RP$^2$), Mobius strip and Klein bottle. The RP$^2$  is a sphere with one crosscap,  the Mobius strip is a disc 
with one crosscap whereas Klein bottle is a sphere with two 
crosscaps. Therefore to construct RP$^2$, we remove a disc from the sphere and 
identify the diametrically opposite points of the resulting hole, whereas
to construct the Mobius strip, we remove a disc from the interior of the disc and 
identify the diametrically opposite points of the resulting hole. A Klein bottle can be 
obtained by removing two discs from the sphere and identifying the diametrically 
opposite points of each of the resulting hole.

We now recall some general statements about two dimensional surfaces. 
Any compact orientable 2 dimensional manifold is topologically equivalent to a 
sphere with $g$ handles and $b$ boundaries. On the other hand, any compact 
non-orientable 2 dimensional manifold is topologically equivalent to a sphere with 
$c$ crosscaps and $b$ boundaries. 
Surfaces with both handles and crosscaps are redundant, since, in the presence 
of crosscaps, a handle can be replaced by 
two crosscaps:
\be
\mbox{handle } \ +\ \mbox{crosscap} \ =\ 3\ \mbox{crosscaps}\, .
\ee
Therefore for non orientable surfaces, the number of boundaries and crosscaps 
specify the surface topologically. 
The Euler number of a general surface, having $g$ handles (or $c$ crosscaps) and $b$ boundaries is given by
\ben\label{etwofor}
\mbox{Orientable surfaces}\qquad &:&\quad\chi = 2-2g-b  \non\\
\mbox{Non orientable surfaces}\qquad &:&\quad \chi = 2-c-b \, .
\een
These determine the power of the string coupling constant $g_s$ in a given amplitude.
Often, one combines the two formula in \refb{etwofor} 
into a single one to write $\chi=2-2g-c-b$ with 
the understanding that we choose $g=0$ for the non-orientable 
surfaces and $c=0$ for the oriented surfaces. 
In the same convention, the dimension of moduli space of an arbitrary two 
dimensional surface with $N$ bulk punctures and $M$ boundary
punctures is given by 
\ben
{\rm dim}\left(\mathcal M_{g,b,c,N,M}\right)\ =\ 6g-6 +3b+3c+2N +M\,.
\een   
Also for getting a non-zero correlation function, the required value of 
the total picture number on a surface is given by
\ben
4g-4+2b +2c\, .
\een
This, together with the picture numbers carried by the string states, 
dictates the number of PCOs one needs to insert on the surface.

The conformal killing groups of sphere and disc without punctures 
are SL(2,$\mathbb C$) and SL(2,$\mathbb R$) respectively. 
There are 3 complex conformal Killing vectors (CKVs) 
on the sphere and 3 real CKVs on the disc. 
Moreover, the volume of these conformal killing groups is infinite. 
This implies that the 1 and 2 point sphere amplitudes 
do not give any contribution to the 1PI effective action.\footnote{It has 
been argued in a recent
paper\cite{1906.06051} that the two point function on the sphere does not vanish. 
This, however, represents the
standard forward contribution to  the S-matrix present in
any quantum field theory, including
string field theory, and is  needed for unitarity of the theory\cite{1604.01783}.
Therefore this does not require us to add a new term in the action. The same comment 
holds for two point function on the disc.}
Similarly, the 1 and 2 point disc amplitudes for external open strings also do not
give any contribution to the 1PI effective action. 
However, the amplitude of 1 closed string on the disc does not vanish since the 
resulting surface -- disc with one bulk puncture -- 
has only one real CKV, generating a finite volume $U(1)$ group. 
The conformal killing group of RP$^2$ is SU(2) which has finite volume. 
Hence, we can also have non zero 1-point  function of closed strings 
on RP$^2$.

The parametrization of the world-sheet with coordinates $(\sigma,\tau)$ 
defines  an orientation to the world-sheet locally. 
The orientation can be reversed by making the following transformation
\ben
\Omega:\qquad \sigma \rightarrow \sigma' =\ell -\sigma\, ,
\qquad\quad \tau \rightarrow \tau'=\tau\label{5.4.1}
\een
 The parameter $\ell$, describing the length of the string in $\sigma$ coordinates, 
 is usually chosen to be $\pi$ for open strings and $2\pi$ for the closed strings. 
 In complex coordinates $z=e^{\tau+i\sigma}$, the operator $\Omega$ acts as
 \ben
 \mbox{Closed strings}&:&\quad z\rightarrow \bar z\, , \quad\qquad \bar z\rightarrow z\, ,\non\\
  \mbox{Open strings}&:&\quad z\rightarrow -\bar z\, , \quad\qquad \bar z\rightarrow -z\, .
 \een
 The operator $\Omega$ is called world-sheet parity operator. 
 Since acting twice with $\Omega$ gives us the original orientation, we have 
 $\Omega^2=1$ and hence the eigenvalues of $\Omega$ are $\pm 1$. 
 
 Orientifold operation typically corresponds to taking a projection by the operator 
 $\Omega$ possibly accompanied by another symmetry transformation acting on the
 world-sheet superconformal field theory. It could for example involve reversing the
 directions of certain space-time coordinates. Let us denote by $W$ this combined 
 operation. We can take $W^2=1$.\footnote{If $W^2$ is not identity, then it must be
 given by some symmetry $U$ of the world-sheet theory that does not involve world-sheet
 parity transformation. We can first define a theory where we take the quotient of the
 world-sheet theory by $U$. This is an ordinary orbifold 
 superconformal field theory describing
 oriented strings. In this theory $U$ acts as identity operator, and therefore $W^2=1$.
 The desired orientifold is now given by the $W$ quotient of the orbifold theory.}
In the theory obtained by quotienting the original theory by $W$, we keep only
those states that carry $W$ eigenvalue $+1$. 
This can be achieved using the projection operator $P$:
\ben
P\ =\ \f{1+W}{2}\, .
\een
Note that the propagator also has GSO projection in all superstring field theories and $L_0=\bar L_0$
projection in the closed string sector, but we do not
display this explicitly.

\begin{table}[t]
% title of Table
\centering 
\begin{tabular}{|C{2.7cm}|C{2.7cm}|C{2.2cm}|C{3.5cm}|C{2.6cm}|}
\hline% \hline
\textbf{Sewing type} &\textbf{degeneration type} &  \textbf{Boundary sewn} 
  &\textbf{Sewing relation} &\textbf{change in topology} \\ 
\hline\hline
\multirow{3}{*}{ \vspace*{.2in}closed-closed}
&\multirow{3}{*}{ \vspace*{.2in}separating} &\multirow{3}{*}{  \vspace*{.4cm} - }&\multirow{3}{*}{ \vspace*{.2in} $zw=q,\ 
z\bar w=q$ } &\multirow{3}{*}{  \vspace*{.2in}- } \\ [.0cm]
&  & \vspace*{-.1cm} & &  \\ [.0cm]
%\hline\hline
\hline\hline
\multirow{3}{*}{ \vspace*{.2in}closed-closed}
&\multirow{3}{*}{ \vspace*{.2in}non separating} & \multirow{3}{*}{\vspace*{.4cm} -}&\multirow{3}{*}{ \vspace*{.2in} $zw=q$ }&\multirow{3}{*}{ \vspace*{.2in}\vbox{\hbox{$g\rightarrow g+1$}\hbox{$c\to c+2$}}} \\ [.0cm]
&  & \vspace*{-.1cm}  & & \\ [.0cm]
\hline\hline
\multirow{3}{*}{ \vspace*{.2in}closed-closed}
&\multirow{3}{*}{ \vspace*{.2in}non separating} &\multirow{3}{*}{  \vspace*{.4cm} - }&\multirow{3}{*}{ \vspace*{.2in} $z\bar w=q$ }&\multirow{3}{*}{ \vspace*{.2in}
\vbox{\hbox{$c\to 2g+2$}\hbox{$c\rightarrow c+2$}}} \\ [.0cm]
&  & \vspace*{-.1cm} & &  \\ [.0cm]
\hline\hline
\multirow{3}{*}{ \vspace*{.2in}open-open}
&\multirow{3}{*}{ \vspace*{.2in}separating} &\multirow{3}{*}{  \vspace*{.4cm} different }&\multirow{3}{*}
{ \vspace*{.2in}  \hskip -.1in
 {\small $zw=-e^{-s},\ z\bar w=e^{-s}$} }&\multirow{3}{*}{ \vspace*{.2in}$b\rightarrow b-1$} \\ [.0cm]
&  & \vspace*{-.1cm} & &  \\ [.0cm]
\hline\hline
\multirow{3}{*}{ \vspace*{.2in}open-open}
&\multirow{3}{*}{ \vspace*{.2in}non separating} &\multirow{3}{*}{  \vspace*{.4cm} same }&\multirow{3}{*}{ \vspace*{.2in} $zw=-e^{-s}$ }&\multirow{3}{*}{ \vspace*{.2in}$b\rightarrow b+1$} \\ [.0cm]
&  & \vspace*{-.1cm} & &  \\ [.0cm]
\hline\hline
\multirow{3}{*}{ \vspace*{.2in}open-open}
&\multirow{3}{*}{ \vspace*{.2in}non separating} &\multirow{3}{*}{  \vspace*{.4cm} same }&\multirow{3}{*}{ \vspace*{.2in} $z\bar w=e^{-s}$ }&\multirow{3}{*}{ \vspace*{.2in}
\vbox{\hbox{$c\to 2g+1$}\hbox{$c\rightarrow c+1$}}} \\ [.0cm]
&  & \vspace*{-.1cm} & &  \\ [.0cm]
\hline\hline
\multirow{3}{*}{ \vspace*{.2in}open-open}
&\multirow{3}{*}{ \vspace*{.2in}non separating} &\multirow{3}{*}{  \vspace*{.4cm} different }&\multirow{3}{*}{ \vspace*{.2in} $z w=-e^{-s}$ }&\multirow{3}{*}{ \vspace*{.37in}{\tiny
$g\rightarrow g+1, \ b\rightarrow b-1$}}\\[.2cm]
& & &&{\tiny $c\to c+2, \ b\rightarrow b-1$} \\ [.0cm]
%&  & \vspace*{-.1cm} &   \\ [.0cm]
\hline\hline
\multirow{3}{*}{ \vspace*{.2in}open-open}
&\multirow{3}{*}{ \vspace*{.2in}non separating} &\multirow{3}{*}{  \vspace*{.4cm} different }&\multirow{3}{*}{ \vspace*{.2in} $z\bar w=e^{-s}$ }&\multirow{3}{*}{ \vspace*{.37in}{\tiny $c\rightarrow 2g+2, \ b\to b-1$}}\\[.2cm]
&&&&{\tiny $c\rightarrow c+2, \ b\rightarrow b-1$} \\ [.0cm]
%&  & \vspace*{-.1cm} &   \\ [.0cm]
\hline
\end{tabular}
\caption{Relation between the topology of the sewed surface to the 
topology of the surface(s) before sewing. 
The last column of some of the rows have two entries. The top entry refers to
the case where the original surface before sewing is oriented, while the bottom
entry refers to the case where the original surface was non-orientable.}
\label{table3}
\end{table}

Let us now review how non-orientable 
 surfaces appear in the unoriented theories. Consider 
a loop diagram. In the intermediate state, we need to sum over all the states. Since we only want to keep the $W=1$ states, the intermediate states must include the projection operator $(1+W)/2$. This corresponds to cutting the world-sheet describing the
propagation of the intermediate state, inserting the projection operator $P$ and then
sewing back the cut-edges.
Then the $1$ part of the projector corresponds to gluing the cut edges with the same orientation -- via the sewing relations \refb{esew1} for
closed strings and \refb{esew2} for the open strings. However, the $W$ 
part of the projector corresponds to first reversing the orientation of one edge and 
then gluing it with the other edge. The corresponding sewing relations are
$z\bar w=q\equiv e^{-s-i\theta}$ for closed strings and $z\bar w=e^{-s}$ for open strings.
This produces a non-orientable surface even if the original surface was
oriented. Note that both, the oriented and the non-orientable 
surface produced this way,
come with weight half.

With this understanding we can now describe
the effect of different types of sewing  in the 
unoriented 
theories. 
One general result that one can infer from this is that  sewing
two different surfaces never produces a
non-orientable surface (unless one of the sewed surface itself is non-orientable). 
The reason 
for this is that while sewing two surfaces $\Sigma$ with $\Sigma'$, we can
consider 
the 
sewing of a fixed surface $\Sigma$ with all possible surfaces $\Sigma'$ having the 
same topology. For sewing with $W$, we just change $w\rightarrow \bar w$ 
without changing the topology of $\Sigma'$. Hence, the new surface obtained 
by $w\rightarrow \bar w$ is already included in the original list of surfaces 
$\Sigma'$. 
So, by sewing with $W$, we do not generate any new surface which 
had not been 
already generated by the sewing with 1. In contrast, for non-separating type
degeneration sewing with $W$ produces non-orientable surfaces from oriented 
surfaces for reasons explained earlier.
We reproduce in table  \ref{table3} the results described in \cite{9708084} 
for different type of sewing and the associated degeneration of the sewed
surface.

\end{document}